\documentclass{article}
\usepackage{graphicx,pdfpages}
\usepackage{amssymb,amsmath,xcolors}
\usepackage{subfigure}   
\usepackage{ws-rv-van}   
\usepackage{geometry}
\usepackage{fancyhdr}
\makeindex
\newcommand{\msun}{M_\odot}
\newcommand*\aap{A\&A}

\newcommand*\apjl{ApJL}

\def\simgreat{\lower2pt\hbox{$\buildrel {\scriptstyle >}
   \over {\scriptstyle\sim}$}}
\def\simless{\lower2pt\hbox{$\buildrel {\scriptstyle <}
   \over {\scriptstyle\sim}$}}
\def\msun{\,{\rm M_\odot}}
\def\ergs{\,{\rm erg\,s^{-1}}}
\def\kms{\,{\rm km\,s^{-1}}}

\begin{document}

            



\begin{titlepage}
   \vspace*{\stretch{1.0}}
   \begin{center}
      \LARGE\textbf{Radio pulsars: testing gravity and detecting gravitational waves}\\
   \vspace*{\stretch{0.1}}

   \Large{Delphine Perrodin$^{1,\star}$, Alberto Sesana$^{2,\dagger}$}

   \vspace*{\stretch{0.05}}
   \small{\it $^1$ INAF - Osservatorio Astronomico di Cagliari, Via della Scienza 5, 09047 Selargius (CA), Italy\\
$^2$School of Physics and Astronomy and Institute of Gravitational Wave Astronomy, University of Birmingham, Edgbaston, Birmingham B15 2TT, United Kingdom}
   \vskip  1.0 truecm
   
   
   
   \end{center}
   \vspace*{\stretch{2.0}}

\begin{flushleft}   
   $^\star$ email: delphine@oa-cagliari.inaf.it\\
   $^\dagger$ email: asesana@star.sr.bham.ac.uk\\ 
\end{flushleft}   
   

\end{titlepage}


\def\simgreat{\lower2pt\hbox{$\buildrel {\scriptstyle >}
   \over {\scriptstyle\sim}$}}
\def\simless{\lower2pt\hbox{$\buildrel {\scriptstyle <}
   \over {\scriptstyle\sim}$}}
\def\msun{\,{\rm M_\odot}}
\def\ergs{\,{\rm erg\,s^{-1}}}
\def\kms{\,{\rm km\,s^{-1}}}

\tableofcontents    

\newpage
\begin{abstract}
\textit{Pulsars are the most stable macroscopic clocks found in nature. Spinning with periods as short as a few milliseconds, their stability can supersede that of the best atomic clocks on Earth over timescales of a few years. Stable clocks are synonymous with precise measurements, which is why pulsars play a role of paramount importance in testing fundamental physics. As a pulsar rotates, the radio beam emitted along its magnetic axis appears to us as pulses because of the lighthouse effect. Thanks to the extreme regularity of the emitted pulses, minuscule disturbances leave particular fingerprints in the times-of-arrival (TOAs) measured on Earth with the technique of pulsar timing. Tiny deviations from the expected TOAs, predicted according to a theoretical timing model based on known physics, can therefore reveal a plethora of interesting new physical effects. Pulsar timing can be used to measure the dynamics of pulsars in compact binaries, thus probing the post-Newtonian expansion of general relativity beyond the weak field regime, while offering unique possibilities of constraining alternative theories of gravity. Additionally, the correlation of TOAs from an ensemble of millisecond pulsars can be exploited to detect low-frequency gravitational waves of astrophysical and cosmological origins. We present a comprehensive review of the many applications of pulsar timing as a probe of gravity, describing in detail the general principles, current applications and results, as well as future prospects.}
\end{abstract}

\section{Introduction to pulsar timing}
\label{intro}

Pulsars are highly-magnetized and fast-rotating neutron stars. In particular, radio pulsars emit beams of radio waves, which, thanks to the lighthouse effect, appear to distant observers as {\it pulses} for every rotation of the pulsar. So far we have discovered more than 2000 pulsars in our own Galaxy and the neighbouring Magellanic Clouds \cite{mhth05}. Of particular interest, millisecond pulsars (MSPs) are pulsars with very short rotation periods (1-30 ms) and are often found in binaries. It is now understood \cite{1991PhR...203....1B} that these pulsars have been spun-up during the {\it recycling} process in which a companion star transfers angular momentum to the neutron star. Their very regular pulsations make them extremely stable clocks. Indeed, through the process of {\it pulsar timing}, which consists in monitoring the times-of-arrival (TOAs) of the pulsars' observed pulses over several years of observations, the rotation period of these pulsars can be estimated to 15 significant figures. The monitoring of MSPs therefore allows us to perform {\it high-precision pulsar timing}, with which we can precisely determine the properties of pulsars and their environment, and study the composition of the interstellar medium between Earth and each pulsar \cite{lk05}.

A newly-discovered pulsar is initially determined by its approximate rotation period $P$, dispersion measure $\rm DM$ (representing the integrated column density of free electrons along the line of sight between pulsar and Earth) and its position in the sky. Through pulsar timing, additional parameters characterizing the pulsar and its environment can be determined. A typical pulsar timing campaign consists of the regular monitoring of TOAs from a known pulsar over several years and with a weekly to monthly cadence. Each pulsar observation is divided into a number of time intervals (sub-integrations) and frequency channels (sub-bands). Since radio pulsars are faint and single pulses are rarely directly observable, it is necessary to integrate ({\it fold}) the radio pulses over many rotations of the pulsar to obtain integrated pulse profiles for each sub-integration and each sub-band. In addition, since the dispersion of the radio signal in the interstellar medium means that the higher-frequency signals arrive at the telescope before the lower-frequency signals, it is necessary to perform the process of {\it de-dispersion} of the radio signals within each sub-band. After {\it folding} and {\it de-dedispersing} the radio signals, topocentric TOAs are obtained by comparing the observed pulse profiles with high signal-to-noise standard profiles obtained from observations of the same pulsar over a time span of several years. The precision of our pulsar timing observations is characterized by the precision of the obtained TOAs (TOA error). Meanwhile, we can calculate expected TOAs based on our best-known models for the pulsar parameters. By subtracting the observed TOAs from the expected TOAs, we obtain {\it timing residuals} that are expected to be scattered around a zero mean, and which are characterized by a root-mean-square (rms) value. An excellent match between timing observations and timing model corresponds to a small rms residual.

Studying the pulsar timing residuals and improving the fitting of pulsar parameters enable us to refine our pulsar models. These models include parameters related to the pulsar's rotation (e.g. the period derivative $\dot{P}$) and orbit (when the pulsar is in a binary), which allow us to test gravity in the strong-field regime. Other parameters describe the dispersion of the radio signal in the interstellar medium as well as its time variations. Finally, and of great interest to Pulsar Timing Arrays (PTAs), we could also find, in the resulting timing residuals, the signature for low-frequency gravitational waves (GWs), such as those emitted by supermassive black hole binaries. In particular, in order to detect a background of low-frequency GWs, PTAs study the correlation of timing residuals for an array of pulsars, which are used as {\it cosmic clocks}. It is therefore crucial for PTAs to use pulsars with very high precision, or equivalently low rms residuals. In order to extract a low-frequency GW signal from the timing residuals, we also need to properly account for both pulsar timing noise, which is most likely related to instabilities in pulsar magnetospheres \cite{lhk+10}, and time variations of the dispersion measure. 

We note that topocentric TOAs, which are measured with Earth's telescopes, are not in an inertial frame. They need to be converted to {\it barycentric} TOAs, as if they were observed at the Solar System Barycentre (SSB). To transform topocentric TOAs  to SSB TOAs, that is to perform the process of {\it barycentric correction}, we need to take several time delays related to Earth's orbit within the Solar System into account. There are also delays due to the pulsar's orbit if the pulsar is in a binary. The SSB TOAs $t_{\rm SSB}$ are related to the topocentric TOAs $t_{\rm topo}$ in this way:
\begin{eqnarray}
t_{\rm SSB} & =  & t_{\rm topo} + 
  t_{\rm clock} - k\times {\rm DM}/f^2 \; \\
      & + & (\Delta_{R, \odot} + \Delta_{S, \odot} + \Delta_{E, \odot}) + (\Delta_{R, \text{bin}} +  \Delta_{S, \text{bin}} + \Delta_{E, \text{bin}}),
\end{eqnarray}
where $t_{\rm clock}$ refers to clock correction terms, $k$ is a constant, $\rm DM$ is the dispersion measure and $f$ is the observing frequency. The dominant term in the barycentric correction is the Roemer delay $\Delta_{R,\odot}$, which is the time delay due to light travel across the Earth's orbit. Second, we have the Shapiro delay $\Delta_{S, \odot}$, which is due to the curved gravitational field of the Sun and planets such as Jupiter. Finally, we have the Einstein relativistic time delay $\Delta_{E, \odot}$, which is due to the time dilation from the motion of the Earth, as well as the gravitational redshift from to the Sun and planets in the Solar System. Additionally, if the pulsar is in a binary, there are equivalent time delays due to the orbit of the pulsar and its companion: $\Delta_{R,\text{bin}}$, $\Delta_{S, \text{bin}}$, and $\Delta_{E, \text{bin}}$. In fact, because of their strong-field dynamics, binary pulsars are extremely interesting for performing tests of strong-field gravity. 

In Section~\ref{sec:tests}, we will review the science and main results in the use of radio pulsars (and pulsar timing techniques) in testing gravity in the strong-field regime. In particular, relativistic binaries such as double neutron star (DNS) binaries provide great laboratories for testing General Relativity (GR), while neutron star - white dwarf (NS-WD) binaries are particularly suitable for tests of alternative theories of gravity. In Section~\ref{sec:gws}, we discuss the science and main results in the use of radio pulsars as `cosmic clocks' for detecting gravitational waves from distant supermassive black hole binaries and the limits already placed on such a background of gravitational waves. In Section~\ref{sec:future}, we discuss future prospects for both tests of strong gravity and gravitational wave detection, especially in light of the Square Kilometre Array (SKA). Finally we summarize our results in Section~\ref{sec:sum}. 

\section{Tests of gravity with radio pulsars} \label{sec:tests}

One hundred years have passed since Einstein presented his theory of gravity known as General Relativity (GR) in 1915. Much progress has been made since then to test the validity of GR. The most stunning confirmations of Einstein's theory include the indirect detection of gravitational waves (GWs) through timing observations of the Hulse-Taylor pulsar \cite{ht75}, and the recent, direct detections of GWs from black hole binaries by the advanced Laser Interferometer Gravitational Observatory (LIGO), as predicted by Einstein \cite{LIGO}. Most of the earlier astrophysical tests of GR were done in the Solar System, which corresponds to the weak-field limit of gravity \cite{w93}, that is a regime where the gravitational potential $\epsilon = G M/(R c^2)$ around a test body of mass $M$ and radius $R$ (where $G$ is the gravitational constant and $c$ is the speed of light) is negligible. GR has thus far passed all tests with flying colours in the weak-field limit \cite{w10,w06}. 
However, the strong-field limit of gravity (where the gravitational potential $\epsilon$ is close to unity) has not been extensively tested, and gravity could possibly deviate from GR in this regime, such as in the environments around compact objects like neutron stars and black holes. We note that while the ``strength" of gravity is usually characterized by the gravitational potential $\epsilon$, a more thorough approach also includes the spacetime curvature $\xi \equiv G M/(R^3 c^2)$ \cite{psaltisLRR}. GR has also passed all tests conducted so far in the strong-field regime, including the recent LIGO observations of black hole binaries \cite{LIGO}. A number of  {\it alternative theories of gravity}, which deviate from GR in the strong-field limit, but not in the weak-field limit which has been extensively tested, have been proposed \cite{de96}. Current tests of gravity seek to better constrain GR and alternative theories of gravity (ruling out some theories in the process), in the absence of any GR violation; or to potentially find deviations from GR in the strong-field limit. Why look for a breakdown of GR if it has thus far passed all tests with flying colours? As we know, GR is not compatible with quantum mechanics and could break down at small scales, such as in the interior of black holes where the concept of a black hole singularity is not physical. In addition, the evolution of the universe cannot be properly described by GR unless one adds the concept of  {\it dark energy}, which could be modelled as a cosmological constant in Einstein's equations. The idea is then that GR is not a complete theory and that by testing gravity in the strong-field limit, we might find deviations from it. 

Pulsars are ideal laboratories for testing GR and alternative theories of gravity. Their environments involve strong gravitational fields ($\epsilon \sim 0.2$ at the surface of a neutron star), and they provide us with much information in the form of extremely regular radio pulses. Pulsar binaries, which involve strong gravitational fields in the vicinity of the neutron star as well as high orbital velocities, are especially interesting for testing gravity, since the orbital dynamics depend on the underlying theory of gravity. Through the fitting of post - Keplerian parameters (see Section \ref{sec:PPK} below) in the pulsar TOAs, the orbital dynamics can be determined and the deformation of spacetime around the pulsar can be constrained \cite{wt81,dt92, ehm06}. Pulsars that are in orbit with a compact object provide even more constraining tests of gravity, especially when the two compact objects are in a close orbit. 
Therefore, by finding systems with companions in closer orbits, we are able to test the limits of GR. In GR, the {\it self-energy} of the neutron star does not affect the orbital dynamics. This is not the case in most alternative theories of gravity, where additional scalar, vector or tensor fields affect the spacetime curvature \cite{w93,w10,w06}. We could therefore observe a breakdown of the predictions of GR in these systems. 

Recent and comprehensive reviews have been published on the topics of: experimental gravity \cite{will+14}; astrophysical tests of gravity \cite{berti}; tests of gravity with radio pulsars \cite{wex}. Recent reviews on tests of gravity with radio pulsars also include \cite{marta, kramer2016}, and \cite{shao2014} discusses in particular all of the ways in which the Square Kilometre Array (SKA) will improve current gravity tests with pulsars. In this section, we outline the methods used to constrain GR and alternative theories of gravity with radio pulsars and present the most important results (best constraints) achieved thus far. Future prospects, in particular with the SKA, will be discussed in section~\ref{sec:future}. The main methods with which radio pulsars can probe gravity involve: the Parametrized Post-Keplerian (PPK) formalism in pulsar binaries, including relativistic spin effects, as discussed in section~\ref{sec:PPK}, and the Parametrized Post-Newtonian (PPN) formalism which quantifies deviations from GR (section~\ref{sec:PPN}). We outline the best constraints on GR using the Double Pulsar in section~\ref{sec:DP} and the best constraints on scalar-tensor theories of gravity (using mostly pulsar - white dwarf binaries) in section~\ref{sec:AT}.

\subsection{Testing gravity with the PPK formalism}
\label{sec:PPK}

In the context of Newtonian physics, binary systems can be described by five Keplerian parameters: the orbital period $P_b$, the orbital eccentricity $e$, the projected semi-major axis $x \equiv a \sin i$, the longitude of periastron $\omega$, and the time of periastron passage $T_0$. The {\it mass function} depends on the Keplerian parameters $P_b$ and $x$:
\begin{equation}
f(M) \equiv \frac{(M_c \sin i)^3}{(M_P+M_c)^2} = \frac{4 \pi^2 x^3}{G P_b^2},
\label{massfunction}
\end{equation}
where $M_P$ is the mass of the pulsar, $M_c$ is the mass of the companion, and $G$ is the gravitational constant. In the context of GR however, we will see below that we also need to include Post-Keplerian (PK) parameters that describe the relativistic effects beyond keplerian orbits, and which constitute excellent tools for testing gravity in binary pulsars.

Since GR is highly non-linear, it does not provide an exact, analytic description of the motion of two bodies. When compact objects move at less than relativistic speeds (the orbital velocity $v/c$ is small), the dynamics of the system can be described by the Post-Newtonian (PN) approximation. In this formalism, the equations of motion are described by a series expansion based on powers of the small parameter $(v/c)^{2n}$, where $n$ is the order of the PN expansion and the 0-th term corresponds to Newtonian dynamics. In fact, the motion of relativistic binaries is adequately described by the PN approximation for most of the binary's inspiral (the orbital velocity is high enough that PN terms are necessary to account for relativistic corrections; however when the velocity is too close to the speed of light right before the merger, the PN expansion breaks down). The 1PN dynamics in binaries -- first order in the PN expansion, which corresponds to terms up to $(v/c)^2$ -- is described by the quasi-Keplerian parametrization of Damour \& Deruelle  \cite{dd85,dd86}. Furthermore, Damour \& Taylor proposed the Parametrized Post-Keplerian (PPK) formalism, which is a phenomenological parametrization based on the quasi-Keplerian parametrization \cite{d88,dt92}: it parametrizes the effects observed in both pulsar timing and pulse structure data. It is theory-independent, which allows us to test both GR and alternative theories of gravity, and consists of a Post-Keplerian (PK) set of parameters that describe the dynamics of relativistic binaries. 

PK parameters are a function of known Keplerian parameters (supposedly already known to high precision), leaving only the two masses as unknowns: the pulsar's mass $M_p$ and the companion's mass $M_c$. Therefore the measurement of two PK parameters leads to the determination of the two masses. By constraining more PK parameters, we can also constrain (or exclude) theories of gravity. $N$ PK parameters will yield $N-2$ tests for any chosen gravity theory. These PK parameters, which are included in pulsar timing models and therefore determined with years of pulsar data (always gaining higher precision with longer data spans), are best plotted in a $M_p$ - $M_c$ diagram. If PK constraints overlap in a mass-mass plot for a particular gravity theory, the particular theory of gravity is still considered a possible valid theory of gravity. If the PK constraints do not overlap, that theory is excluded \cite{marta,wex}.

In GR, the most important PK parameters are: the variations of two Keplerian parameters $\omega$ and $P_b$ defined in section~\ref{intro}, i.e. the relativistic precession of periastron $\dot{\omega}$ and the change in the orbital period due to the back-reaction of gravitational wave emission on the binary motion $\dot{P_b}$. Additionally, we have the time delays such as the Einstein delay related to the changing time dilation of the pulsar clock (due to variations in orbital velocity) and gravitational redshift $\gamma$, and the range $r$ and shape $s$ of the Shapiro delay related to a changing gravitational redshift in the gravitational field of the companion \cite{s64}. 
Their expressions as a function of the Keplerian parameters $P_b$, $x$, $e$ and the two masses $M_p$ and $M_c$ are shown below \cite{dd86,tw89,dt92}:
\begin{eqnarray}
\dot{\omega} &=& 3 \, T_\odot^{2/3} \left( \frac{P_{\rm b}}{2\pi} \right)^{-5/3} \,
               \frac{1}{1-e^2} \; (M_p + M_c)^{2/3}, \label{equ:omegadot}\\
\gamma  &=& T_\odot^{2/3} \left( \frac{P_{\rm b}}{2\pi} \right)^{1/3} \,
              e \, \frac{M_c(M_p+2M_c)}{(M_p+M_c)^{4/3}}, \\
r &=& T_\odot \, M_c, \\
s & \equiv & \sin i \, =
  T_\odot^{-1/3} \left( \frac{P_{\rm b}}{2\pi} \right)^{-2/3} \; x \;
           \frac{(M_p+M_c)^{2/3}}{M_c}, \\\dot{P}_{\rm b} &=& -\frac{192\pi}{5} T_\odot^{5/3}  \left( \frac{P_{\rm b}}{2\pi} \right)^{-5/3} 
            \frac{\left(1+\frac{73}{24} e^2+ \frac{37}{96} e^4\right)}{(1-e^2)^{7/2}}
            \frac{M_p M_c}{(M_p + M_c)^{1/3}}, \label{equ:pbdot}
\end{eqnarray} 
where masses are expressed in solar units, $T_\odot \equiv G M_{\odot}/c^3 = 4.925490947 \mu \rm s$, $G$ is Newton's gravitational constant and $c$ is the speed of light. Additional PK parameters of interest include the change in orbital eccentricity $\dot{e}$ and the change in the projected semi-major axis $\dot{x}$. The relativistic precession of periastron $\dot{\omega}$ is easiest to measure in eccentric orbits, while the Shapiro parameters $r$ and $s$ are measurable in nearly edge-on binary systems. In alternative theories of gravity, the expressions for the PK parameters are slightly different and include theory-dependent parameters that can be constrained \cite{w93,w06}.

\subsubsection{Double neutron star binaries} \label{sec:double}

The first real test of gravity in the strong-field regime was accomplished by Hulse and Taylor in 1974 with the discovery of PSR B1913+16 (dubbed the {\it Hulse-Taylor pulsar}), which was the first {\it binary pulsar} ever discovered in the radio band. It consists of a pulsar in a double neutron star (DNS) binary \cite{ht75}. The measurement of two PK parameters ($\dot{\omega}$ and $\gamma$) enabled the precise determination of the two neutron star masses (assuming GR was correct) \cite{wnt10}. Having fully determined the binary system, any additional test would constitute a test of GR. In fact, the measurement of the decrease in the orbital period $\dot{P}$, associated with a loss of orbital energy, was found to be consistent with GR's predictions \cite{tfm79}. Specifically, it is consistent with GR's {\it quadrupole formula} that describes the backreaction of GW emission on the binary motion \cite{1963PhRv..131..435P}. 
This confirmed GR's predictions and provided the first indirect detection of GWs as predicted by Einstein. The agreement between the measured $\dot{P_b}$ and the predicted GR value is currently at the 0.2 \% level \cite{wnt10}.

The pulsar PSR J0737-3039, discovered at Parkes in 2003 \cite{bdp+03,lbk+04} is, like the Hulse-Taylor pulsar, composed of a DNS binary. In addition, the second neutron star has been observed as a pulsar; this system is therefore dubbed {\it the Double Pulsar} with two pulsars: PSR J0737-3039A ({\rm psr A} with a period of 22 ms) and PSR J0737-3039B ({\rm psr B} with a period of 2.7s). The Double Pulsar is a profoundly unique system for testing gravity, since the radio pulses from both stars provide two clocks that can be monitored with pulsar timing. It is also characterized by large orbital velocities and a closeness of the orbit, which both amplify the importance of relativistic effects, and the high orbital inclination makes its timing easier. In this system, five PK parameters have been determined: $\dot{\omega}$, $\gamma$, $r$ and $s$ and $\dot{P_b}$ \cite{bdp+03}. Additionally, the sizes of both pulsars' orbits were estimated and the mass ratio $R$, which is independent of the theory of gravity, was measured for the first time in a DNS system \cite{lbk+04}. 

DNS binaries are ideal systems for testing GR. In recent years, an increasing number of DNS systems have been discovered: so far, more than 15 DNS systems are known \cite{2017Tauris}. 
In the next few years, more pulsar surveys (in particular with the SKA, see Section \ref{sec:future}) will discover new DNS binaries and further constrain GR.

\subsubsection{Relativistic spin effects} \label{sec:spin}

Not all relativistic effects can be described at the 1PN level with the PK parameters. For example, tests of relativistic gravity can be done at 2PN \cite{de96} or 2.5 PN \cite{mw13}. 
In addition, binary pulsars can have spin. The spin terms appear at higher orders in the Post-Newtonian expansion \cite{bo75, d87, p06, b91}. In particular, the presence of spin-orbit coupling terms (the coupling of the spin of one pulsar with the binary's angular momentum) in the binary's equations of motion leads to the {\it Lense-Thirring precession of the orbit} or  {\it frame-dragging}, as well as a change in the projected semi-major axis $\dot{x}$ \cite{bo75,1988NCimB.101..127D, sp09}. Additionally, time-dependent spin terms in the equations of motion lead to changes in the orientations of the pulsar spins (which we refer to as {\it relativistic spin precession} or {\it geodetic precession}) \cite{dr74, bo75, ber75}. The precession of the pulsar's rotation axis is essentially being caused by the curvature of spacetime from the companion star. This effect can be seen in changes in the pulsar emission: changes in the spin axis of the pulsar makes different regions of the magnetosphere visible to the observer, thus affecting the observed pulse profile.


In the Double Pulsar, the contribution to the Lense-Thirring precession is dominated by the fast-rotating {\rm psr A}. However, $\dot{x}$ is difficult to measure because of the near alignment of pulsar spin and orbital angular momentum \cite{f13}. Future measurements of the Lense-Thirring precession with the Double Pulsar is discussed in \cite{kehl2016}. The Double Pulsar is however the best system we know so far for testing relativistic spin precession \cite{sp09}. Indeed, the relativistic precession of {\rm psr B}'s spin axis can be determined thanks to the eclipses of {\rm psr A} (that is when {\rm psr A} passes behind {\rm psr B}). Its precession rate was measured and found to be compatible with GR with an uncertainty of 13 \%: $\Omega_B = (4.77\pm^{0.66}_{0.65})^\circ \, \text{yr}^{-1}$ \cite{bkk+08,pmk+10}. Relativistic spin precession has also been observed in the following binary pulsars: PSR B1913+16 \cite{k98, wt02,cw08}, PSR B1534+12 \cite{sta04, fst14}, J1141-6545 \cite{mks+10} and J1906+0746 \cite{lsf+06}. J0737-3039B and PSR B1534+12 are the only two pulsars for which we have a direct measurement of the precession rate (and which matches GR predictions) \cite{sta04,fst14,bkk+08,pmk+10}. 

\subsubsection{Best test of GR: The Double Pulsar}
\label{sec:DP}

\begin{figure}[h!]
\centerline{\includegraphics[width=0.6\textwidth, angle=0]{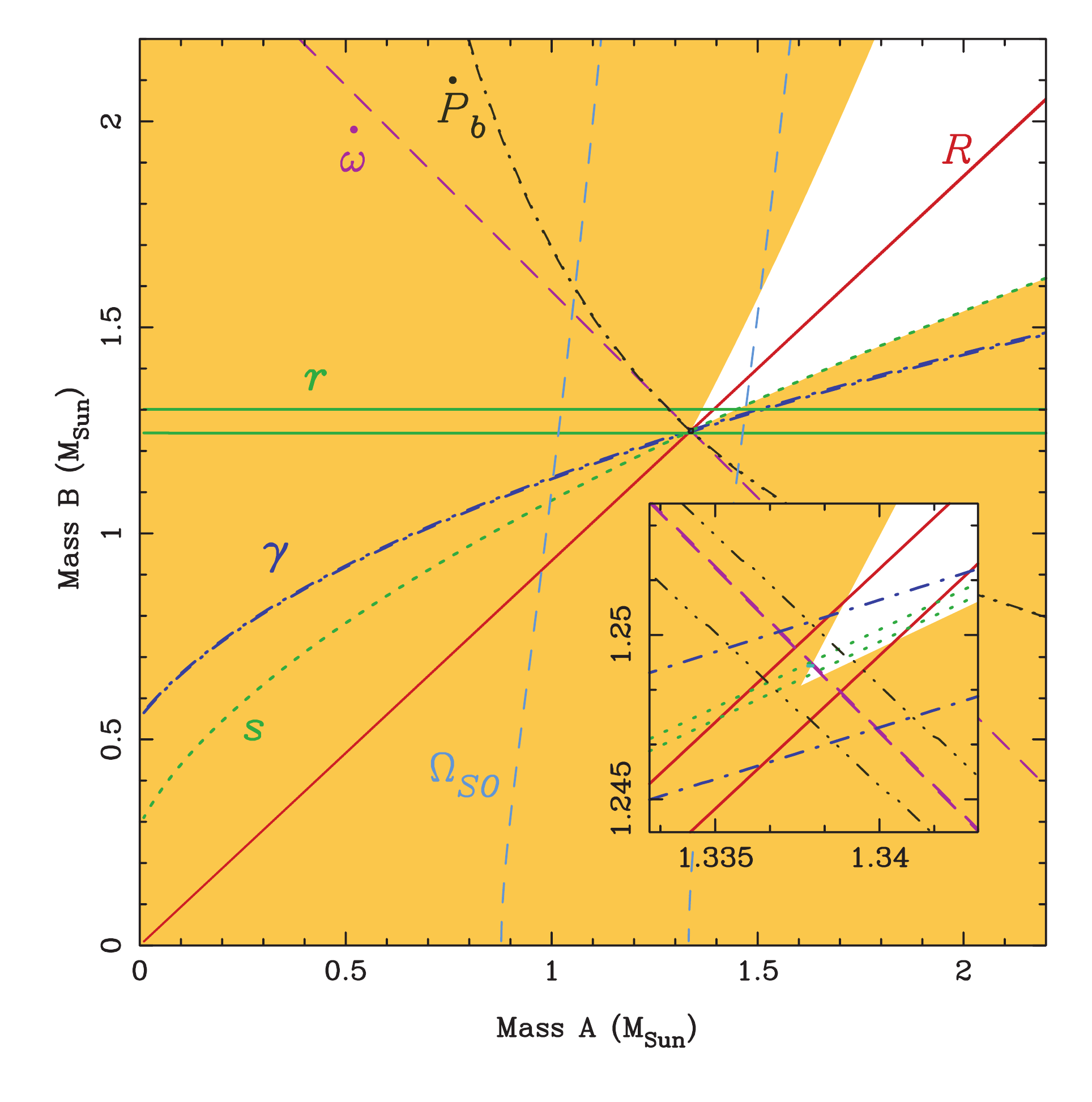}}
\caption{Mass-mass diagram for the Double Pulsar J0737-3039. Shaded regions are excluded by the Newtonian mass functions (one for each pulsar). The five PK parameters ($\dot{P_b}$, $\dot{\omega}$, $\gamma$, $r$ and $s$), the mass ratio $R$ and the precession rate of {\rm psr B} ($\Omega_{\rm SO}$) constrain the remaining parameter space, providing multiple tests of GR. So far GR is verified with an uncertainty of 0.05 \% (figure courtesy of Michael Kramer).}
\label{fig:DP}
\end{figure}

In the Double Pulsar, we have a total of seven mass constraints, thanks to the determination of five PK parameters, the mass ratio $R$ (see section~\ref{sec:double}),
and the precession rate $\Omega_B$ \cite{ks08} (see section~\ref{sec:spin}). In addition, there are constraints related to the Newtonian mass function, one for each pulsar (see equation~(\ref{massfunction})). The masses of both pulsars are determined with high precision, leaving us with an additional five tests of GR, as shown in Fig.~\ref{fig:DP}. The Double Pulsar provides thus far the most stringent test of GR, with an uncertainty of 0.05 \% \cite{k+06}. The longer we continue to monitor this system, the more precise the TOAs, and the better the GR constraints we will obtain. In particular, $\dot{\omega}$ could be determined up to the 2PN order, and the spin of {\rm psr A} could be determined. With an even better determination of $\dot{P_b}$ (such as that expected thanks to the interferometric determination of the parallax \cite{dbt09}), the Double Pulsar will also provide stringent constraints on alternative theories of gravity that predict the presence of dipolar gravitational radiation. Through a measurement of its moment of inertia \cite{1988NCimB.101..127D}, the Double Pulsar could also constrain the equation of state of nuclear matter in neutron star interiors \cite{watts_ska}.

\subsection{Testing gravity using the PPN formalism}
\label{sec:PPN}

As we have seen in the previous sections, the fitting of PK parameters in the timing data of pulsar binaries allows us to determine the masses of the binary companions (if at least two PK parameters are measured) and to constrain gravity theories (if more than two PK parameters are measured). In addition, the study of the variations in pulse profiles allows us to determine changes in the spin precession of pulsars. These tools can also be applied to test the Strong Equivalence Principle (SEP), Lorentz invariance or conservation of momentum. The SEP is unique to GR: any violation of the SEP is a violation of GR \cite{w93}. The SEP has been tested extensively in the Solar System, that is in the weak-field limit, using the PPN formalism \cite{w10, wn72, w93, w06}. We refer the reader to \cite{w06} for a full description of the formalism. The main idea is that in any metric theory of gravity, the dynamics (i.e. the equations of motion) of objects in a gravitational field depends exclusively on the structure of the metric. Therefore, any measurable departure from GR for a given theory has to be characterized by some difference in its metric compared to the GR one. In the weak-field limit, the most general metric can be written as an expansion of the Minkowski spacetime with the addition of ten (small) PPN parameters. Pulsars can test the SEP using the same formalism, providing in this way complementary tests to Solar System tests, since they can test gravity (GR and alternative theories of gravity) in the strong-field limit \cite{s03}. This however requires a modification of the original ten PPN parameters to account for strong-field effects \cite{de96, de92} (the original PPN expansion is valid in the weak-field limit). The information we collect from pulsars with the determination of PK parameters can be translated into constraints on PPN parameters (PK parameters describe small variations in the motion of compact binaries, which can be mapped into small variations of the underlying metric). The ten (modified) PPN parameters describe the existence of preferred frames, preferred locations, the non-conservation of momentum, the non-linear superposition of gravitational effects, or the space-time curvature produced by a unit mass (for a full definition and physical interpretation of each individual parameter, see \cite{w06,marta}).

 \subsubsection{SEP violation and orbital dynamics}

The SEP includes both the Weak Equivalence Principle (WEP) and the Einstein Equivalence Principle (EEP). The WEP tests the universality of free fall, stating that the trajectory of a free-falling body in a gravitational field should be independent of its internal structure. A first test of the SEP can therefore be accomplished by comparing the trajectories of two massive objects in a gravitational field, for example by looking for a {\it polarization} in the direction of the gravitational potential (this is the Nordtvedt effect or {\it gravitational Stark effect}, \cite{n68}. Lunar Laser Ranging (LLR) experiments have tested the Nordtvedt effect by comparing the Earth and the Moon's free falls in the Sun's gravitational potential, and have imposed strong constraints on PPN parameters for the Solar System \cite{LLR}. Similarly, we can look at the two companions of a pulsar binary and how they {\it fall} in the gravitational potential of the Galaxy. It works best if the two companions are different in mass and composition, therefore double neutron star binaries (DNS) are not ideal laboratories for testing SEP violations. Instead, a sample of pulsars with white dwarf companions (PSR-WD) can impose strong constraints on SEP violations \cite{sfl+05, gsf+11,freire2012}, in particular on the following parameter:
\begin{equation}
\Delta = \left( \frac{M_{\rm grav}}{M_{\rm inertial}} \right)_1 - \left(\frac{M_{\rm grav}}{M_{\rm inertial}}\right)_2,
\end{equation}
where $M_{\rm grav}$ is the gravitational mass and $M_{\rm inertial}$ is the inertial mass of each body. So far the best constraint on $\Delta$ is from a study of 27 PSR-WD binaries \cite{gsf+11}: $\Delta < 4.6 \times 10^{-3}$ (see Table~\ref{Table1}).

The discovery of an MSP (PSR J0337+1715) in a triple system with two white dwarf companions \cite{rsa+14} will allow us to greatly improve the constraint on the SEP. The masses of the three bodies have all been determined. The two inner masses (the pulsar and inner WD), of different masses and composition, are moving in the gravitational field of the outer WD, which is larger than that of the Galaxy by at least six orders of magnitude, therefore the SEP violation would be greatly magnified. This system could therefore be the best laboratory we have so far to constrain the SEP, with an estimated constraint on the parameter $\Delta$ of four orders of magnitude better than current constraints, most likely with the use of future telescopes \cite{rsa+14, shao2014, berti, s+16}.


\subsubsection{SEP violation: violation of LLI and LPI}

The EEP states that local, non-gravitational experiments are independent of the frame. The EEP consists of the Local Lorentz Invariance (LLI) and Local Position Invariance (LPI). Violations of LLI correspond to the observation of a preferred frame, while violations of LPI correspond to the observation of preferred positions, 
and may also lead to variations in fundamental constants such as the gravitational constant $G$. Violations of LLI and LPI both involve changes in the orbital dynamics of binary pulsars and the spin precession of solitary pulsars, which are characterized by PK parameters such as the changes in orbit eccentricity $\dot{e}$, inclination $\dot{x}$, and the periastron advance rate $\dot{\omega}$. Testing of LPI can in particular be done by looking at the spin precession of pulsars: a violation of LPI could be seen if we observe changes in the expected pulsar spin precession around the acceleration toward the galactic centre. This would be evident by studying the stability of the pulse profiles of solitary pulsars. 

The violations of LLI and LPI, which, for binary pulsars, are determined by changes in the aforementioned PK parameters, are characterized by the following PPN parameters: $\hat{\alpha}_1$, $\hat{\alpha}_2$ and $\hat{\alpha}_3$, where the $\hat{}$ refers to the strong-field generalization of the associated PPN parameter. The parameters $\hat{\alpha}_1$ and $\hat{\alpha}_2$ involve the existence of a preferred frame (i.e. non-zero values would imply a violation of LLI). $\hat{\alpha}_2$ also includes the spin precession of the pulsar, which can be seen from changes in pulse profiles. A non-zero value of the PPN parameter $\hat{\alpha}_3$ involves both the existence of a preferred frame (a violation of LLI) and a violation of conservation of momentum \cite{w93}. 
The parameter $\hat{\xi}$, which is the strong-field equivalent of the Whitehead PPN parameter $\xi$, characterizes LPI violation through measurements of the spin precession; a limit on $\hat{\xi}$ can be converted into a constraint on the spatial anisotropy of the gravitational constant $G$ \cite{sw13}. The parameter $\hat{\zeta_2}$ characterizes non-conservation of momentum through the measurements of the polarization of the orbit and the spin precession.  Additionally, the SEP would be violated if gravitational dipole radiation is observed. This would represent an obvious violation of GR, and the constraints on proposed alternative theories of gravity would become fundamental. 

We find that the timing analysis of the PSR-WD binary PSR J1738+0333 leads to some of the best constraints on PPN parameters. Additionally, it is also the best pulsar so far to constrain scalar-tensor gravity (see Section~\ref{sec:AT}). Other interesting and complementary constraints are obtained from the pulse profile analysis of isolated MSPs PSR B1937+21 and PSR J1744-1134. The best constraints on $\hat{\alpha}_1$, $\hat{\alpha}_2$, $\hat{\alpha}_3$, $\hat{\xi}$ and $\hat{\zeta}_2$ are listed in Table~\ref{Table1}. 

\subsubsection{Varying gravitational constant}

Violation of LPI can lead to variations in fundamental constants such as the gravitational constant $G$. PSR J0437-4715 is one of the best pulsars for high precision timing because of its closeness to Earth and its brightness \cite{jlh93,dtb+09,k95,k96,sbb+01}. The inclination angle can be determined independently of the theory of gravity and compared to the expected Shapiro delay ($s= \sin i$). They are in good agreement. Until recently, this pulsar provided the best test of $\dot{G}$ using pulsar binaries: $| \frac{\dot{G}}{G} | <  23 \times 10^{-12} \, \text{yr}^{-1}$ \cite{vbs+08}. 
Recent measurements of PSR J1713+0747 however show a straighter constraint: $| \frac{\dot{G}}{G}|  <  (-0.6 \pm  1.1) \times 10^{-12} \, \text{yr}^{-1}$  at 95\% CL \cite{weiwei}. This is the best limit on $\dot{G}/G$ using pulsar binaries. 

\begin{table}[h!!!]
\centering
\begin{tabular}{ | c | c | l |}
\hline 
Parameter & Upper limit & Method \\ [0.5ex]
\hline \hline
$\Delta$ & $5.6 \times 10^{-3}$ (95\% CL) & PSR-WD binaries \cite{sfl+05} \\  & $4.6 \times 10^{-3}$ (95\% CL) & PSR-WD binaries \cite{gsf+11} (see \cite{wex} for discussion) \\ [0.8ex]
\hline
$\hat{\alpha}_1$ & $\left(-0.4^{+3.7}_{-3.1}\right) \times 10^{-5}$ (95\% CL) & timing analysis of PSR J1738+0333  \\ &  & better than solar system \cite{sw12,avk+12, fwe+12} \\ 
\hline
$\hat{\alpha}_2$ & $1.6 \times 10^{-9}$ (95\% CL) & timing analysis of PSR J1738+0333 \\ & & + pulse profile data of B1937+21/J1744-1134 \\ &  & better than solar system \cite{avk+12, fwe+12,sck13} \\
\hline
$\hat{\alpha}_3 $& $4 \times 10^{-20}$ (95\% CL) & PSR-WD binaries (better than solar system) \cite{sfl+05} \\ 
\hline
$\hat{\xi}$ & $3.9 \times 10^{-9}$ (95\% CL) & pulse profile data of B1937+21/J1744-1134 \\ & & better than solar system \cite{sw13} \\
\hline
$| \frac{\Delta G}{G} |^{\rm anis.} $ &  $4 \times 10^{-16}$  & derived from $\hat{\xi}$ constraint \cite{sw13} \\ [0.8ex]
\hline
$\hat{\zeta_2}$ & $4 \times 10^{-5}$ & non-conservation of momentum from B1913+16 \cite{w92} \\
\hline
$| \frac{\dot{G}}{G} | $ &  $[(-0.6 \pm  1.1)] \times 10^{-12} \, \text{yr}^{-1}$ (95\% CL) & J1713+0747 \cite{weiwei} \\ [0.8ex]
\hline
dipolar & 0.002 (95\% CL) & J1738+0333 \cite{fwe+12} \\
$| \alpha_{\rm P} - \alpha_0 | $ & 0.005 (95\% CL) & J0348+0432 \cite{afw+13} interesting because of massive NS \\
\hline
\end{tabular}
\caption{Best constraints on PPN parameters characterizing deviations from the Strong Equivalence Principle (SEP); on the spatial anisotropy $|\Delta G/G|$ and time variation $|\dot{G}/G|$ of the gravitational constant; on dipolar radiation via the parametrization $| \alpha_{\rm P} - \alpha_0 |$.}
\label{Table1}
\end{table}


\subsection{Tests of alternative theories of gravity}
\label{sec:AT}

Pulsars allow us to test both GR and alternative theories of gravity: we may either detect a {\it breakdown} of GR; or confirm GR and place limits on alternative theories of gravity, such as scalar-tensor theories. In the particular case of {\it tensor-mono-scalar theories} \cite{de93,de96b}, gravity is mediated by the metric field $g_{\mu\nu}$ as well as a scalar field $\phi$. These theories are characterized by the following coupling between matter and the scalar field $\phi$:
\begin{equation}
a(\phi) = \alpha_0 \phi + 1/2 \beta_0 \phi^2
\end{equation}
This formalism includes GR in the case where $\alpha_0 = \beta_0 = 0$. It also includes the Jordan-Fierz-Brans-Dicke theory \cite{bd61,wz89} in the case where $\beta_0=0$ and $\alpha_0 = \sqrt{1/(2 \omega_{BD} + 3)}$, where $\omega_{BD}$ is the Brans-Dicke parameter. 
Variations in the scalar field $\phi$ could produce observable effects such as a gravitational constant varying with space and time (non-zero $\dot{G}$) \cite{de92,w93} or the detection of gravitational dipole radiation in a pulsar binary, either of which would constitute a violation of the SEP and a {\it breakdown} of GR. The existence of a varying gravitational constant or dipole gravitational radiation would affect the PK parameters, most particularly the orbital decay $\dot{P_b}$ \cite{d88,wex}. In the absence of an obvious breakdown of GR (no detection of $\dot{G}$ or GW dipole radiation), the $(\alpha_0, \beta_0)$ parameter space can be constrained by binary pulsar observations \cite{fwe+12}. We note that non-perturbative effects such as spontaneous scalarization could also affect the dynamics of the binary system \cite{de93}. While the DNS systems such as B1913+16, B1534+12 and J0737-3039 provide constraints on scalar-tensor theories, they are not the best sources for testing alternative theories of gravity such as scalar-tensor gravity. Indeed, in the case of two identical neutron stars, the dipolar gravitational radiation term essentially vanishes. Pulsars with WD companions, with different masses and compositions, can better constrain these theories. In most PSR-WD binaries, only two PK parameters can be determined (the Shapiro delay parameters $r$ and $s$), allowing a determination of the two binary masses, however that is not enough for constraining gravity theories \cite{avk+12,afw+13,wex}. Interestingly, the following PSR-WD binaries allow for the determination of more than 2 PK parameters: J1141-6545, J1738+0333, J0437-4715 and J0348+0432. They provide tests that are complementary to the GR tests using DNS J0737-3039 and B1913+16 \cite{marta}. 

\begin{itemize}

\item{In PSR J1141-6545, three PK parameters can be determined: $\dot{\omega}$, $\gamma$, and $\dot{P_b}$ \cite{klm+00a}. This has led to the determination of both masses and one test of GR at the 10\% level \cite{bbv08}. The pulsar's relativistic spin precession can also be observed \cite{mks+10}, but is not as well measured as for the Double Pulsar or B1913+16. It is however useful for constraining scalar-tensor theories and possibly detecting dipolar gravitational radiation.}

\item{PSR 1738+0333 is so far the most useful pulsar for constraining scalar tensor theories \cite{jhb+05,fwe+12,wex}, as it provides a precise determination of $\dot{P_b}$ (in good agreement with GR), as well as proper motion and parallax, giving the best upper limit on dipolar GW (see Table~\ref{Table1}). }

\item{PSR J0348+0432, discovered in 2013 \cite{b+13,l+13}, has the highest mass of any pulsar observed so far: $2.01 \pm  0.04 \, M_{\odot}$. It provides a stringent constraint on $\dot{P_b}$, which is currently at the 82\% agreement with GR, leading to a constraint on dipolar GW radiation, though its upper limit is not as high as J1738+0333 (see Table~\ref{Table1}). Spontaneous scalarization in such a massive system creates an important amount of gravitational dipolar radiation, which rules out an important part of the parameter space in alternative theories; this pulsar also places constraints on a long-range field \cite{afw+13}. Finally, thanks to its high mass, J0348+0432 constrains the equation of state of nuclear matter, favouring a stiff equation of state \cite{watts_ska}.}

\end{itemize}

\begin{figure}[h!!!!!!!]
\centerline{\includegraphics[width=0.8\textwidth]{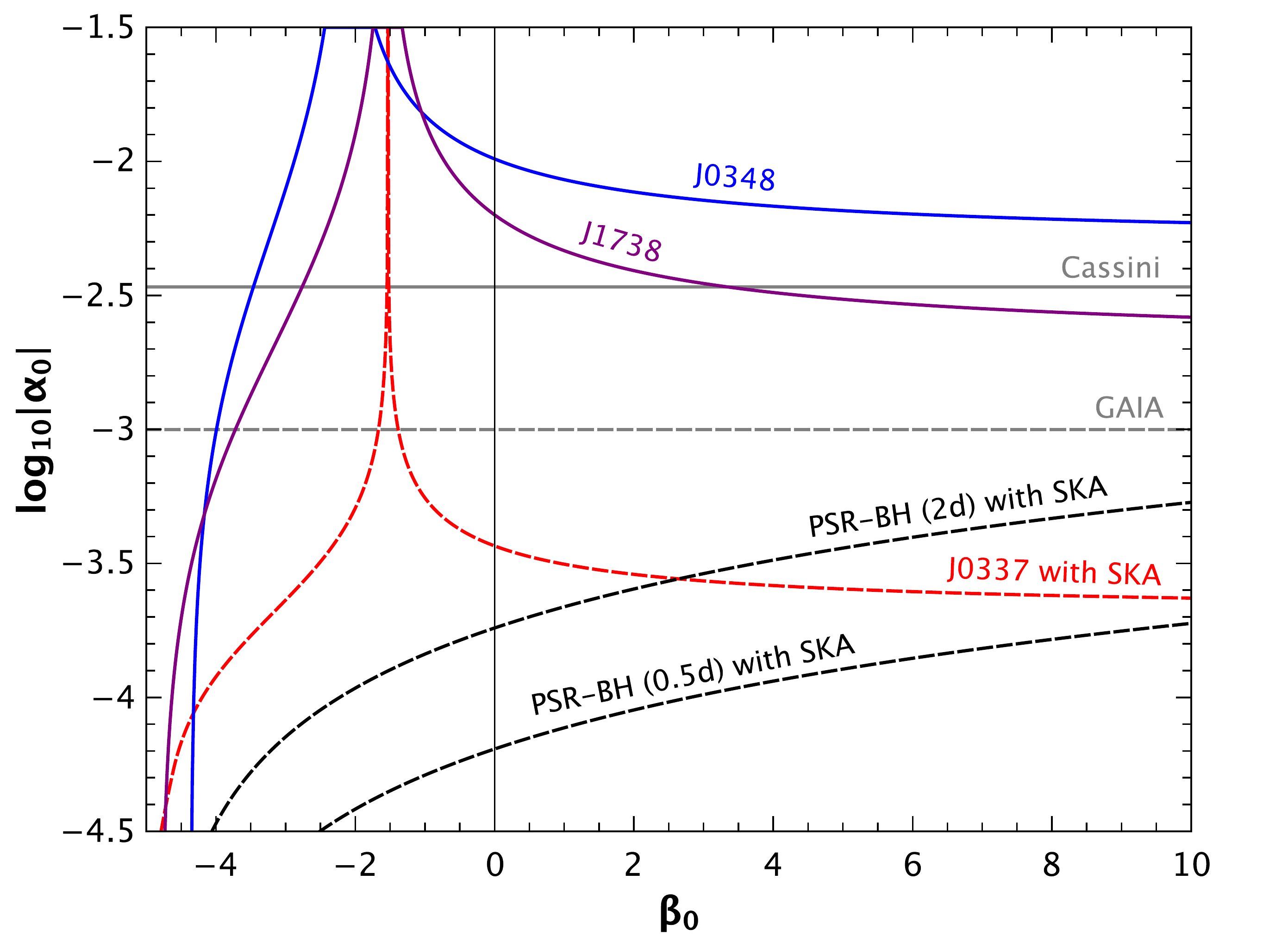}}
\caption{Constraints on the coupling parameters $\alpha_0$ and $\beta_0$ in tensor-mono-scalar theories with coupling $\alpha(\phi) = \alpha_0 \phi + 1/2 \beta_0 \phi^2$ (figure courtesy of Norbert Wex). GR is located at the intersection of $\alpha_0=0$ and $\beta_0=0$, while Jordan-Brans-Dicke theories, for which $\beta_0=0$, are along the y-axis. The allowed parameter space is constrained to the area below all of the solid lines. The PSR-WD binaries J1738+0333 (purple solid line) and PSR J0348+0342 (blue solid line) provide the best constraints so far, and are comparable to Solar System constraints such as with the Cassini spacecraft (grey solid line) and the future GAIA astrometric satellite (grey dashed line). The triple system PSR J0337+1715 (red dashed line) will likely impose stronger constraints in the near future, especially as observed with the SKA. In addition, we show the expected constraints (black dashed lines) from two hypothetical PSR-BH systems we expect to find with the SKA (with orbital periods $P_b = 2 \rm d$ and $P_b=0.5 \rm d$, respectively). We see that the pulsar timing of PSR-WD systems (such as the triple system PSR J0373+1715) and PSR-BH systems are complementary: the former imposes strong constraints at positive $\beta_0$'s, while the latter imposes strong constraints at negative $\beta_0$'s. These estimates are based on a stiff NS equation of state (MPA1), making the constraints rather conservative \cite{kramer2016, shao2017}.}
\label{Fig:TS}
\end{figure}

So far, PSR J1738+0333 and PSR J0348+0432 provide the best constraints on scalar-tensor gravity theories (including Jordan-Brans-Dicke theory for which $\beta_0=0$); their constraints are comparable to solar system tests such as the Cassini probe (see Fig.~\ref{Fig:TS}) \cite{fwe+12}.  They also provide the best constraints on quadratic scalar-tensor gravity (for  for $\beta_0 < -3$ and $\beta_0 >0$ \cite{fwe+12,berti, kramer2016,wex}. J1738+0333 also excludes TeVeS-like theories \cite{fwe+12}. Massive Brans-Dicke theories are best constrained by PSR J1141-6545 \cite{abw+12}, while Einstein-Aether theories are best constrained by a combination of pulsars: the PSR-WD binaries J1141-6545, J1738+0333, J0348+0432 together with the Double Pulsar J0737-3039 \cite{ybby+14}. We note that the triple system PSR J0337+1715 will likely impose even stronger constraints in the near future \cite{rsa+14, s+16,berti}. The discovery of a pulsar - black hole (PSR-BH) system would also further constrain the parameter space of scalar-tensor theories \cite{liu+12,liu+14,wex+13}. 


\section{Gravitational wave detection with radio pulsars}
\label{sec:gws}

Because of the exquisite stability of MSPs, pulse TOAs are extremely sensitive to any type of perturbation affecting the photon path from the source to Earth, such as variations in the interstellar medium (ISM), solar wind, etc. (see section \ref{sec:noise} below). This makes MSPs formidable tools for detecting GWs. In fact, the passage of a GW between a pulsar and the Earth modifies the null geodesic along which the photons propagate, resulting in small alterations of the pulse TOAs. This was realized even before the discovery of the first MSP \cite{1978SvA....22...36S,1979ApJ...234.1100D}, by applying the mathematical formalism developed by Estabrook and Wahlquist \cite{1975GReGr...6..439E} for detecting GWs using Doppler spacecraft tracking to pulsars. Early work based on a handful of regular pulsars made use of the technique to constrain a putative low-frequency GW background (GWB) of cosmic origin to the level of about $\Omega_{\rm gw}\approx10^{-4}$ times the critical density of the Universe \cite{1983ApJ...265L..39H,1983MNRAS.203..945B,1983ApJ...265L..35R}. In particular, \cite{1983ApJ...265L..39H} proposed that the effect of a GWB is encoded in the peculiar correlation of TOAs collected from pairs of pulsars at different sky locations, and worked out the analytical form of the pattern, which is now known as the Hellings \& Downs curve and is at the heart of current GWB searches with PTAs. The idea was elaborated by Foster \& Backer \cite{1990ApJ...361..300F}, who proposed the concept of a Pulsar Timing Array (PTA), consisting in the regular monitoring of a number of the newly-discovered MSPs \cite{1982Natur.300..615B}. By just monitoring two MSPs, \cite{1994ApJ...428..713K} improved the limit on a stochastic GWB to $\Omega_{\rm gw}=6\times10^{-8}$. In the early 2000s, three major collaborations formed with the goal of providing systematic timing residuals on a sizable ensemble of MSPs: the European Pulsar Timing Array (EPTA \cite{2016MNRAS.458.3341D}), the Parkes Pulsar Timing Array (PPTA \cite{2016MNRAS.455.1751R}) and the North American Nanohertz Observatory for Gravitational Waves (NANOGrav, \cite{2015ApJ...813...65T}). The three collaborations also share data under the aegis of the International Pulsar Timing Array (IPTA, \cite{2016MNRAS.458.1267V}), with the goal of obtaining a combined, more sensitive dataset. Altogether, the three PTAs are timing approximately fifty of the best MSPs with a weekly cadence ($\Delta{t}$) and for a timespan $T$ of several years (more than 20 in some cases), with a timing precision ranging from a few microseconds to a few tens of nanoseconds. PTAs are therefore sensitive to GWs in the frequency range $1/T<f<1/(2\Delta{t})$, corresponding to a few to a few hundred nanohertz. Putative GW signals in this frequency range include those from cosmological stochastic backgrounds from inflation, phase transitions or cosmic strings \cite{2016PhRvX...6a1035L}, but the loudest GW source is expected to be the cosmic population of inspiralling supermassive black hole binaries (SMBHBs), formed following galaxy mergers \cite{2008MNRAS.390..192S}.


%

\subsection{Detection principle}
\label{sec:principles}
To elucidate the detection principle of PTAs, we follow the derivation in \cite{Maggiore2017}. Let us consider a pulsar $p$ pulsating regularly as a perfect clock. A modification in the photon path will result in the pulses arriving slightly earlier or later. The net result is therefore a change in the pulsation frequency $\nu(t)$ observed on Earth, i.e. a redshift (or Doppler shift):
\begin{equation}
  z(t) = \frac{\nu(t)-\nu_0}{\nu_0} = \frac{\delta \nu(t)}{\nu_0},
\label{redshift}
\end{equation}
where $\nu_0$ is the intrinsic pulsar frequency. To establish the potential of PTAs as GW detectors, we need to compute the redshift that is induced by a GW crossing the line of sight to the pulsar. As an analogy with spacecraft Doppler tracking studied by \cite{1975GReGr...6..439E}, it can be demonstrated that in a conformal flat spacetime, for a wave $h_{ij}(t)$ incident on a pulsar located in direction $\hat{p}$, the observed redshift at time $t$ is
\begin{equation}
  z(t) = \frac{1}{2}p_ip_j\int_{t_p}^{t}dt'\frac{\partial}{\partial{t'}}h_{ij}[t', (t-t_p)\hat{p}].
  \label{zte}
\end{equation}
Let us now take, without any loss of generality, the case of a wave incident in the $z$ direction $\hat\Omega=(0,0,-1)$, and a pulsar located in the $(x,z)$ plane in direction $\hat{p}=(\sin\theta_p, 0, \cos\theta_p)$, so that $\pi-\theta_p$ is the angle between the direction to the pulsar and the direction of the incoming wave. We restrict our discussion to GR, so that the wave only has tensor components identified by $h_{xx}=-h_{yy}$, $h_{xy}=h_{yx}$. In this case, we have $n_in_j=\sin^2\theta$ and, after some manipulations, equation (\ref{zte}) gives
\begin{equation}
  z(t) = \frac{1}{2}(1-\cos\theta)[h_{xx}(t_p)-h_{xx}(t)].
  \label{ztesingle}
\end{equation}
Here $t-t_p\equiv\tau=(L/c)(1+\hat\Omega\cdot\hat{p})$ is the difference in TOAs of the incident wave at the pulsar and at the Earth, where $L$ is the Earth-pulsar distance. We note that the redshift $z$ is given by the difference between the metric perturbations at the pulsar at the time of radio emission $t_p$ and the metric perturbations at the Earth at the time of observation $t$. Equation (\ref{ztesingle}) provides some insight about the response of a pulsar to an incoming wave. If the GW source and the pulsar are located on opposite sides (as seen from Earth), then $\theta=0$ and the resulting redshift vanishes. If, on the other hand, the GW source is located right behind the pulsar, then $t=t_p$ and the two metric perturbation components cancel exactly, again giving zero redshift. This is consistent with the transverse nature of GWs.  

The actual quantity measured in PTA experiments is the timing residual $r(t)$. This is simply given by the integral over observing time of the redshift induced by the incident GW:
\begin{equation}
  r(t) =\int_0^t dt'z(t',\hat\Omega),
  \label{res}
\end{equation}
where $t$ is the time of a given pulsar observation, and the integral starts from the beginning of the timing experiment.

\subsubsection{generalization of the residual formula}
\label{sec:resgeneral}
Equation (\ref{ztesingle}) describes the response of the pulsar-Earth detector to an incoming wave in the $z$ direction. It is useful to generalize the formula in two ways. First, although for any given source-pulsar pair we can always define a frame in which the source is in the $z$ direction and the pulsar lies in the $(x,z)$ plane, PTAs combine observations of an ensemble of pulsars \cite{1990ApJ...361..300F}. It is therefore useful to write the pulsar response in a generic fixed frame which does not have a specific alignment with respect to the source-Earth-pulsar reference. Second, equation (\ref{ztesingle}) is expressed in terms of the GW component along the direction defined by the projection of the pulsar location into a plane perpendicular to the incident wavefront (direction $x$ in this case). It is however useful to write the response in terms of the two tensor polarizations of the GW wave $h_{+}$ and $h_{\times}$.

We consider a Cartesian reference frame $(x,y,z)$ centred at the solar system barycentre\footnote{As discussed in section \ref{intro}, TOAs are computed by converting the pulse arrival time at the observatory to the pulse arrival time at the solar system barycentre. In fact, when we refer to 'TOAs measured on Earth',  'GW Earth term' etc., those have to be intended 'at the solar system barycentre'.}. The source $\hat\Omega$ and pulsar $\hat{p}$ locations are therefore defined in terms of the standard angles $(\theta,\phi)$:
\begin{subequations}
\begin{align}
\hat\Omega &= 
- (\sin\theta \cos\phi)\, \hat{x}
- (\sin\theta \sin\phi)\, \hat{y}
- \cos\theta \hat{z}\,
\label{sourcecoor}
\\
\hat{p} &= 
 (\sin\theta_p \cos\phi_p)\, \hat{x}
 + (\sin\theta_p \sin\phi_p)\, \hat{y}
 + \cos\theta_p \hat{z}.
\label{pulsarcoor}
\end{align}
\end{subequations}
Note the minus sign in $\hat\Omega$, which is defined as the direction of the {\it incoming} wave. 

The wave propagating from the $\hat\Omega$ direction consists of two polarization states $h_{+}$ and $h_{\times}$. The relation between those and the metric perturbation along a specific direction is given by
\begin{equation}
h_{ij}(t,\hat\Omega) = e_{ij}^+(\hat\Omega)h_+(t,\hat\Omega) + 
e_{ij}^{\times}(\hat\Omega)h_\times(t,\hat\Omega),
\label{hij}
\end{equation}
where the polarization tensors $ e_{ij}^A(\hat\Omega)$ (with $A=+,\times$) are defined as
\begin{subequations}
\begin{align}
e_{ij}^+(\hat\Omega) &=  \hat{m}_i \hat{m}_j - \hat{n}_i \hat{n}_j\,,
\label{e:e+}
\\
e_{ij}^{\times}(\hat\Omega) &= \hat{m}_i \hat{n}_j + \hat{n}_i \hat{m}_j\,.
\label{e:ex}
\end{align}
\end{subequations}
Here $\hat{m}$, $\hat{n}$ are the GW principal axes and define, together with the direction of the wave propagation $\hat\Omega$, an orthonormal basis. Note that $\hat{m}$ is aligned with the plus wave polarization. We therefore have two Cartesian coordinate systems: one is the `detector frame' defined by $(\hat{x},\hat{y},\hat{z})$, and one is the `wave propagation frame' defined by $(\hat{m},\hat{n},\hat\Omega)$. To compute the response in the detector frame, one needs to project onto it the metric perturbation defined along the principal axes $\hat{m}$, $\hat{n}$ of the wave propagation frame. The principal axis $\vec{m}$ defines an angle $\psi$ (counter-clockwise about the wave propagation) with the line of nodes of the detector frame. We can thus perform a rotation by an angle $\psi$ to express $\hat{m}$, $\hat{n}$ in the detector frame coordinates \cite{2001PhRvD..63d2003A}:
\begin{subequations}
\begin{align}
\hat{m} & = 
(\sin\phi \cos\psi - \sin\psi \cos\phi \cos\theta) \hat{x}  - (\cos\phi \cos\psi + \sin\psi \sin\phi \cos\theta) \hat{y} + (\sin\psi \sin\theta) \hat{z}\,,\\
\hat{n} & = 
(-\sin\phi \sin\psi - \cos\psi \cos\phi \cos\theta) \hat{x}  + (\cos\phi \sin\psi - \cos\psi \sin\phi \cos\theta) \hat{y} + (\cos\psi \sin\theta) \hat{z}.
\end{align}
\end{subequations}

Now that we have defined all of the relevant quantities with respect to the detector frame, equation (\ref{ztesingle}) can be generalized to
\begin{equation}
z(t,\hat\Omega)  =  \frac{1}{2} \frac{\hat{p}^i\hat{p}^j}{1+\hat{p}^i\hat\Omega_i}\left\{e_{ij}^+(\hat\Omega)\left[h_+(t_p,\hat\Omega)-h_+(t,\hat\Omega)\right]-e_{ij}^\times(\hat\Omega)\left[h_\times(t_p,\hat\Omega)-h_\times(t,\hat\Omega)\right]\right\},
\label{ztgeneral}
\end{equation}
which can be written in compact form as 
\begin{equation}
z(t,\hat\Omega) = \sum_A F^A(\hat\Omega)[h_{A}(t_p,\hat\Omega)-h_{A}(t,\hat\Omega)]
\label{e:z1}
\end{equation}
where
\begin{equation}
F^A(\hat\Omega) = \frac{1}{2} \frac{\hat{p}^i\hat{p}^j}{1+\hat{p}^i\hat\Omega_i} e_{ij}^A(\hat\Omega).
\label{e:FA}
\end{equation}
In practice, this notation separates the physics of GW emission, enclosed in the $h_A$ terms, from all of the geometric factors arising from the transformation between the radiation and the detector frames, which are absorbed in the $F^A$ response functions. Note that the latter are universal, i.e. they do not depend on the nature of the GW signal\footnote{This is true so long as only GR tensor polarizations are considered. In alternative theories of gravity, scalar and vector polarizations might also arise, and require different response functions $F$ \cite{2013LRR....16....9Y}.}. The explicit form of the response functions (or antenna beam patterns) is given by 
\begin{subequations}
\begin{align}
F^+(\hat\Omega) & = \frac{1}{2} \frac{(\hat{m} \cdot \hat{p})^2 - (\hat{n} \cdot \hat{p})^2}{1 + \hat\Omega \cdot \hat{p}}\,,\\
F^\times(\hat\Omega) & = \frac{(\hat{m} \cdot \hat{p})\,(\hat{n} \cdot \hat{p})}{1 + \hat\Omega \cdot \hat{p}}\,.
\end{align}
\end{subequations}
Note that the response functions depend only upon the  three direction cosines $\hat{m} \cdot \hat{p}$, $\hat{n} \cdot \hat{p}$ and $\hat\Omega \cdot \hat{p}$ and are independent of the specific choice of Cartesian detector frame, as expected.

\subsubsection{stochastic background}
The set of equations presented in Section \ref{sec:resgeneral} forms a useful method for computing the redshift (and the associated residual through equation (\ref{res})) induced by an incident deterministic GW with a generic form $h(t)$. We now generalize the derivation for a stochastic GWB generated by the incoherent superposition of uncorrelated sources randomly distributed in the sky. In this case, equation (\ref{e:z1}) is generalized to represent the incoming GWs in the Fourier domain as $\tilde{h}_A(f)$ and by integrating over all possible frequencies and incoming directions to obtain 
\begin{equation}
  z(t)=\sum_{A=+,\times}\int_{-\infty}^\infty df\int d^2\hat{\Omega}\,F^A(\hat{\Omega})\tilde{h}_A(f,\hat{\Omega})e^{-2\pi i f t}\left[1-e^{-2\pi i f \tau}\right],
\label{stochred}
\end{equation}
where $\tau=t-t_p$ has been defined in Section \ref{sec:principles}. The interesting quantity for a GWB is the ensemble average over the stochastic variable $\tilde{h}_A(f)$. Under the assumption of an isotropic stationary and unpolarized background, this ensemble average takes the form \cite{2000PhR...331..283M}
\begin{equation}
  \langle \tilde{h}^*_A(f,\hat{\Omega}) \tilde{h}_A'(fi,\hat{\Omega}') \rangle=\delta(f-f')\frac{\delta^2(\hat\Omega,\hat\Omega')}{4\pi}\delta_{AA'}\frac{1}{2}S_h(f),
\label{ensembleh}
\end{equation}
where $S_h(f)$ is the power spectral density of the GWB, and the $1/2$ factor comes from considering only positive frequencies, i.e. $0<f<\infty$. The ensemble average of the timing residuals observed in a pair of MSPs denoted as $a$ and $b$ then becomes
\begin{equation}
  \langle z_a(t)z_b(t) \rangle= \frac{1}{2}\int_{-\infty}^\infty df S_h(f)\int d^2\hat{\Omega} \frac{1}{4\pi}\sum_{A=+,\times}F_a^A(\hat{\Omega})F_b^A(\hat{\Omega}).
\label{crossredshift}
\end{equation}
The above result comes from substituting equation (\ref{ensembleh}) into equation (\ref{stochred}) and by noticing that all of the terms involving $e^{-2\pi i f \tau}$ can be neglected in the short wavelength limit, which is appropriate for PTAs. PTAs are in fact sensitive to nHz GWs, corresponding to parsec wavelengths, which is much shorter than the distance to the closest known MSP of about 150 pc (typical MSP distances are in the kpc range).

The integral over sky orientations in equation (\ref{crossredshift}) was first computed by \cite{1983ApJ...265L..39H} and takes the form
\begin{eqnarray}
C(\zeta_{ab})=\frac{1}{4}\left[1+\frac{\cos\zeta_{ab}}{3}+4(1-\cos\zeta_{ab})\ln\left(\sin\frac{\zeta_{ab}}{2}\right) \right]\, .
\label{e:hdcurve}
\end{eqnarray}
where $\zeta_{ab}$ is the angle between the pulsars $a$ and $b$ on the sky. Finally, the observable quantity in PTA observations is the ensemble average cross correlation in the timing residuals between two pulsars $r_{ab}=\langle r_a(t)r_b(t)\rangle$, where $r_x(t)$ is defined by equation (\ref{res}). We can therefore integrate equation (\ref{crossredshift}) over time to get the final form of the correlation in the timing residuals
\begin{equation}
  r_{ab}=C(\zeta_{ab})\int_0^\infty df \frac{S_h(f)}{4\pi^2 f^2}
\label{e:rab}
\end{equation}
Elaborating on equation (\ref{ensembleh}), one can define a dimensionless characteristic strain $h_c(f)$ satisfying the relation \cite{2000PhR...331..283M} 
\begin{equation}
  h^2_c(f)=2fS(f).
\label{e:hcdef}
\end{equation}
Note that with this definition, $h_c$ is connected to the energy density $\rho_{\rm GW}$ of the GWB via
\begin{equation}
\label{eq:omegagw}
 \frac{2\pi^2}{3H_0^2}f^2 h^2_c(f)=\frac{1}{\rho_c}\frac{\mathrm{d} \rho_{\mathrm{gw}}}{\mathrm{d}\ln f}\equiv\Omega_{\mathrm{gw}}(f),
\end{equation}
where $\rho_c=3H_0^2/8\pi$ is the critical energy density of a flat  Universe, and $H_0=100~h$~km~s$^{-1}$~Mpc$^{-1}$ is the Hubble expansion rate. Substituting equation (\ref{e:hcdef}) into equation (\ref{e:rab}), we finally get
\begin{equation}
  r_{ab}=\Gamma(\zeta_{ab})\int_0^\infty df P_h(f),
\label{e:rabPh}
\end{equation}
where we defined 
\begin{equation}
  P_h(f)=\frac{h^2_c(f)}{12\pi^2f^3},
\label{e:Ph}
\end{equation}
and we re-defined the Hellings \& Downs (HD) correlation coefficients as
\begin{equation}
  \Gamma(\zeta_{ab})=\frac{3}{2}C(\zeta_{ab})(1+\delta_{ab}).
\label{eq:HD} 
\end{equation}
Note that $P_h(f)$ has dimensions of $[s^{-3}]$, which is appropriate for a spectral density of a time series. The $3/2$ renormalization and the $\delta_{ab}$ term ensure that the new correlation coefficient $\Gamma(\zeta_{ab})=1$ when $a=b$ (i.e., the GWB has perfect autocorrelation). Note that when $a \neq b$ and $\zeta_{ab}=0$, $\Gamma_{ab}=1/2$; this is because for pulsars at the same location, but at different distances, only the Earth terms are phase correlated, whereas the pulsar terms act as an additional source of noise. We will see below that this has important implications for GWB detection with PTAs.   

\subsection{GW sources relevant to PTAs and their signals}
In Section \ref{sec:principles} we demonstrated that both deterministic and stochastic GW sources affect the pulse TOAs. Deterministic sources leave a distinctive fingerprint of the form $r(t)$ (cf equation (\ref{res})) that can be exactly determined once the waveform $h(t)$ is known. On the other hand, stochastic GWBs induce a correlated signal $r_{ab}$ (cf equation (\ref{e:rabPh})) that can be determined if the characteristic strain spectrum $h_c(f)$ of the GWB is known. We now discuss the GW sources relevant to PTAs and their signals. 

\subsubsection{Deterministic GW signals}
A signal is deterministic when its waveform can be univocally specified at any given time, pending the knowledge of the signal dependence on the physical parameters of its source. Most of the expected deterministic signals in the PTA band are related individual SMBHBs \cite{1980Natur.287..307B}, inspiralling and merging along the cosmic history (see e.g., \cite{2003ApJ...582..559V}), although more exotic sources have been proposed, such as (super)strings cusps and kinks \cite{2001PhRvD..64f4008D}. Deterministic GW signals can be either continuous or transient; we will see below that SMBHBs can produce either type of signals depending on their physical properties and in which stage of their evolution they are observed.

\vspace{0.3cm}
\hspace{-0.4cm}{\bf I - Inspiralling supermassive black hole binaries}\\
The archetypal continuous deterministic GW source is a SMBHB adiabatically inspiralling in a quasi-circular orbit. PTAs are sensitive to systems with $M>10^{8}\msun$ at centi-parsec orbital separations \cite{2009MNRAS.394.2255S,2012MNRAS.420..860S}. For those systems, the inspiral time is typically much longer than the observation time $T$. In the circular orbit approximation, the system emits a monochromatic wave at twice its orbital frequency (i.e. $\omega=2\omega_K$) of the form \cite{2010PhRvD..81j4008S}:

\begin{subequations}
\begin{align}
h_+(t) & = (1 + \cos^2 \iota) A \cos(\omega t+\Phi_0)\,,
\label{e:hpluscross1}
\\
h_{\times}(t) &=-2 \cos\iota \, A \sin(\omega t+\Phi_0)\,,
\label{e:hpluscross2}
\end{align}
\end{subequations}
where
\begin{equation}
A(f) = 2 \frac{{\cal M}^{5/3}}{D_l}\,\left(\pi f\right)^{2/3}\approx 1.3\times 10^{-15}\left(\frac{f}{10^{-7}{\rm Hz}}\right)^{2/3}\left(\frac{\cal M}{10^9\msun}\right)^{5/3}\left(\frac{D_l}{1{\rm Gpc}}\right)^{-1}
\label{e:Agw}
\end{equation}
is the GW amplitude, $D_l=(1+z)D_c$ the luminosity distance to the GW source, ${\cal M}=(M_1M_2)^{3/5}/(M_1+M_2)^{1/5}$ is the chirp mass (being $M_1$ and $M_2$ the masses of the two SMBHs), $\iota$ is the inclination of the SMBHB orbital plane with respect to the line-of-sight and $\Phi(t) = 2\pi\int^t f(t') dt'$ is the GW phase, being $f=\omega/(2\pi)$. Note that equation (\ref{e:Agw}) is written in terms of redshifted quantities. Those are related to their binary-rest frame counterparts via $\mathcal{M} =(1+z)\mathcal{M}_{{\rm rf}}$, $f=f_{\rm rf}/(1+z)$.\footnote{Unless otherwise specified, we always use {\it redshifted} masses and frequencies to describe the GW signals.} In general, the GW community prefers redshifted quantities because they are the direct observables of GW experiments, and because they absorb all $(1+z)$ factors, simplifying the equation when dealing with sources at cosmological distances. 

The associated redshift $z(t)$ can be computed by plugging equation (\ref{e:hpluscross1},\ref{e:hpluscross2}) into equation (\ref{e:z1}). By integrating the redshift according to equation (\ref{res}), the residual is found to be composed of a pulsar and an Earth term:
\begin{equation}
r(t) = r^p(t)-r^e(t), 
\end{equation}
where
\begin{eqnarray}
r^e(t)  = \frac{A}{\omega} \left\{(1 + \cos^2{\iota}) F^{+} \left[ \sin(\omega t + \Phi_0)  - \sin{\Phi_0}\right] + \right.\nonumber \\
\left. 
2 \cos{\iota} F^{\times} \left[ \cos(\omega t + \Phi_0)  - \cos{\Phi_0}\right]
\right\},\nonumber \\
r^p(t)  = \frac{A_p}{\omega} \left\{
(1 + \cos^2{\iota}) F^{+} \left[ \sin(\omega_p t + \Phi_p + \Phi_0)  - \right.\right. 
\left.\left.  \sin({\Phi_p +\Phi_0})\right] + \right.\nonumber \\
\left.  2 \cos{\iota} F^{\times} \left[ \cos(\omega_p t + \Phi_p + \Phi_0)  -     \right. \right. 
\left. \left.  \cos(\Phi_p + \Phi_0)\right] \right\}.\nonumber \\
\label{Eq:signal}
\end{eqnarray}
We specify $\omega$ and $\omega_{p}=\omega(t-\tau)$, because the GW frequency might be different in the pulsar and Earth terms, also implying different amplitudes ($A$ and $A_p$). In fact, in the quadrupole approximation, the evolution of the binary orbital frequency and GW phase can be written as
\begin{equation}
\omega_K(t)=\omega_K\left(1-\frac{256}{5}{\mathcal{M}}^{5/3}\omega_K^{8/3}t\right)^{-3/8},
\label{Eq:freqev}
\end{equation}
\begin{equation}
\Phi(t)=\Phi_0+\frac{1}{16{\mathcal{M}}^{5/3}}\left(\omega_K^{-5/3}-\omega_K(t)^{-5/3}\right).
\label{Eq:phaseev}
\end{equation}
Over the typical PTA experiment duration (decades),  $\omega$ and $\omega_{p}$ can be approximated as constants, and we drop the time dependence accordingly. However, the delay $\tau$ between the pulsar and the Earth term, which is comparable with the pulsar-Earth light travel time, is, over thousands of years, comparable with the evolution timescale of typical SMBHBs \cite{2010PhRvD..81j4008S}.
$\omega_p$ depends on the pulsar distance and relative orientation with respect to the incoming GW source; it is therefore different among observed pulsars and is smaller than $\omega$.

\begin{figure*}
\centering
\begin{tabular}{cc}
  \includegraphics[width=5.9cm,clip=true,angle=0]{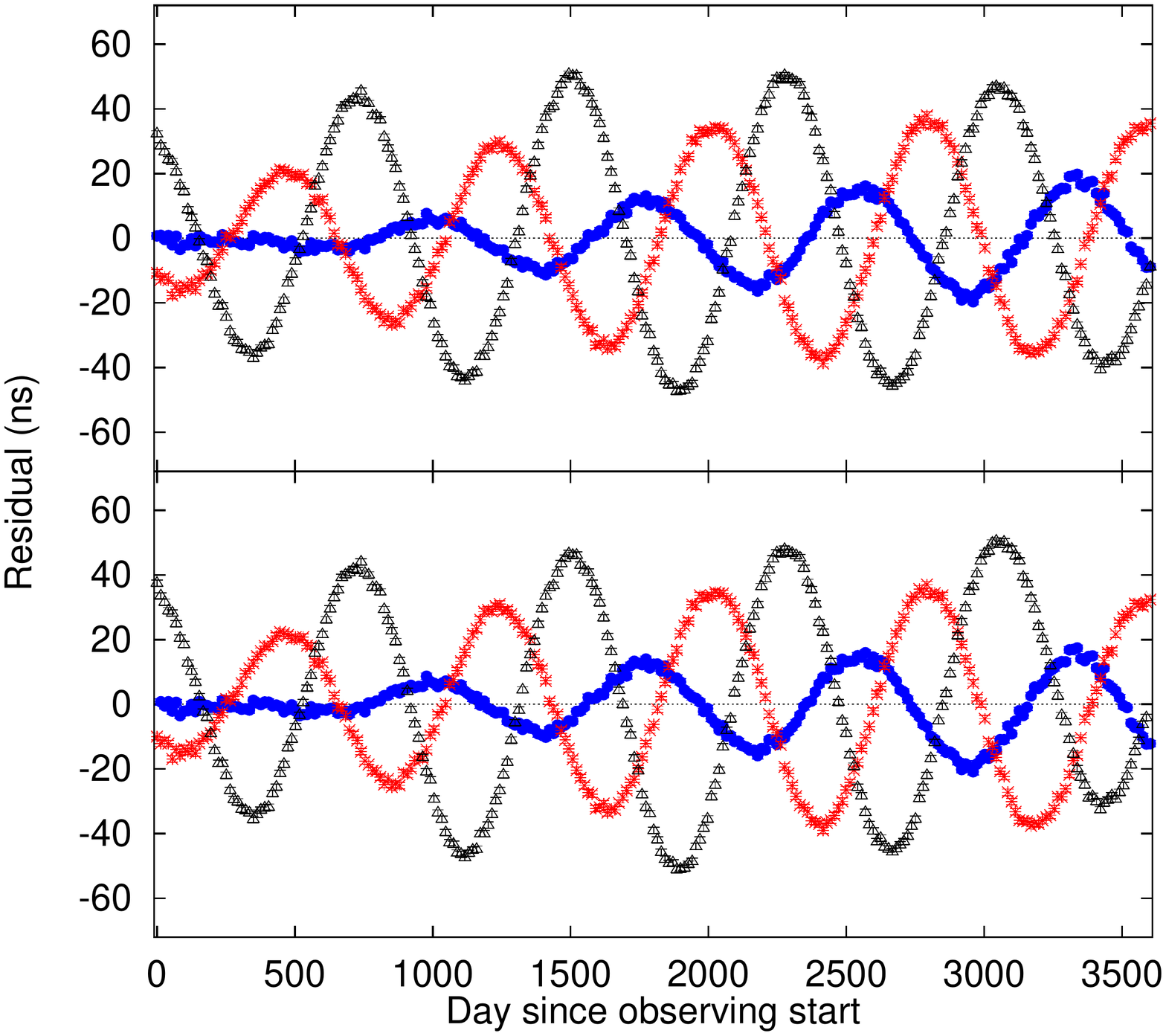}&
  \includegraphics[width=5.9cm,clip=true,angle=0]{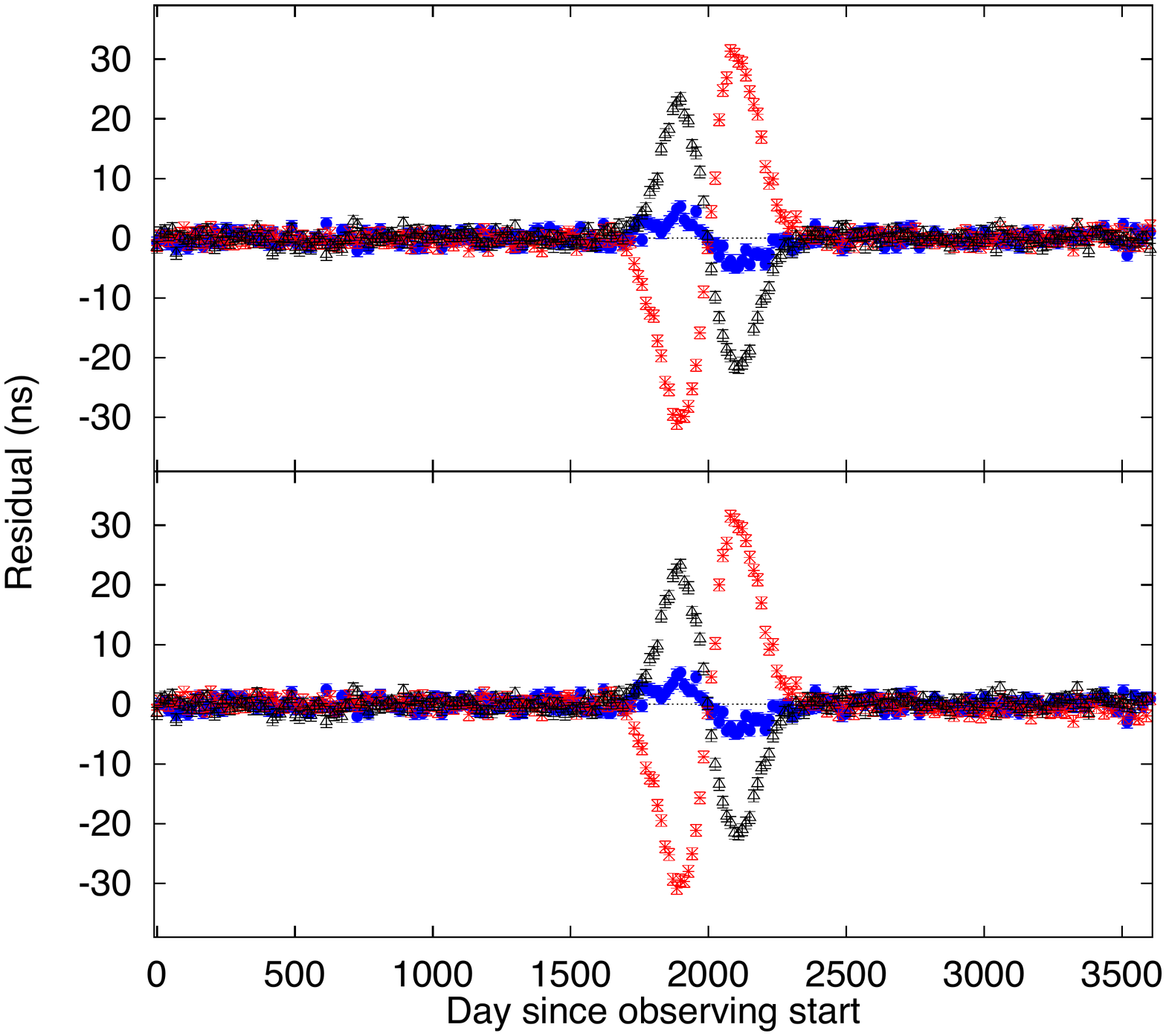}\\
  \includegraphics[width=5.9cm,clip=true,angle=0]{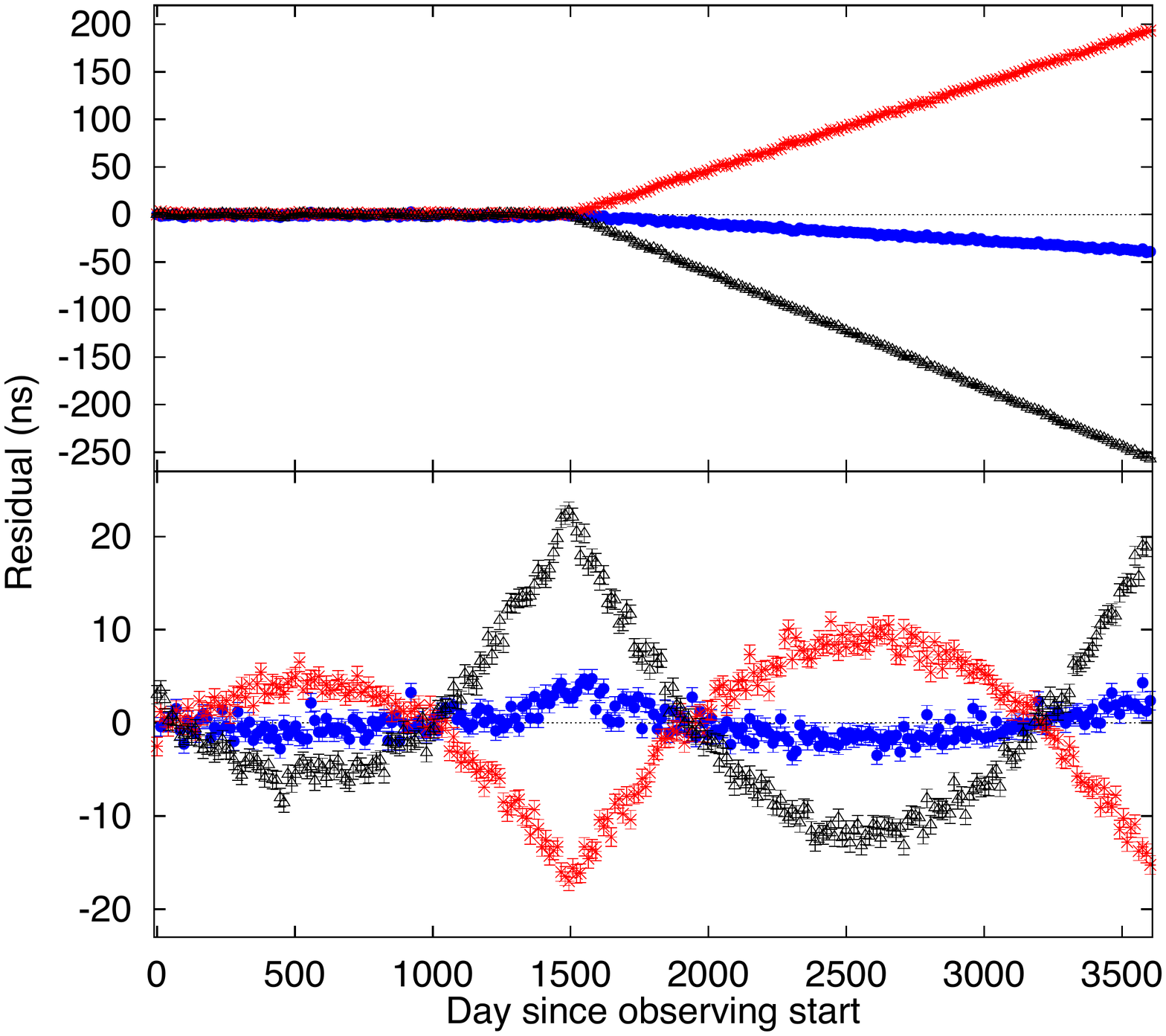}&
  \includegraphics[width=5.9cm,clip=true,angle=0]{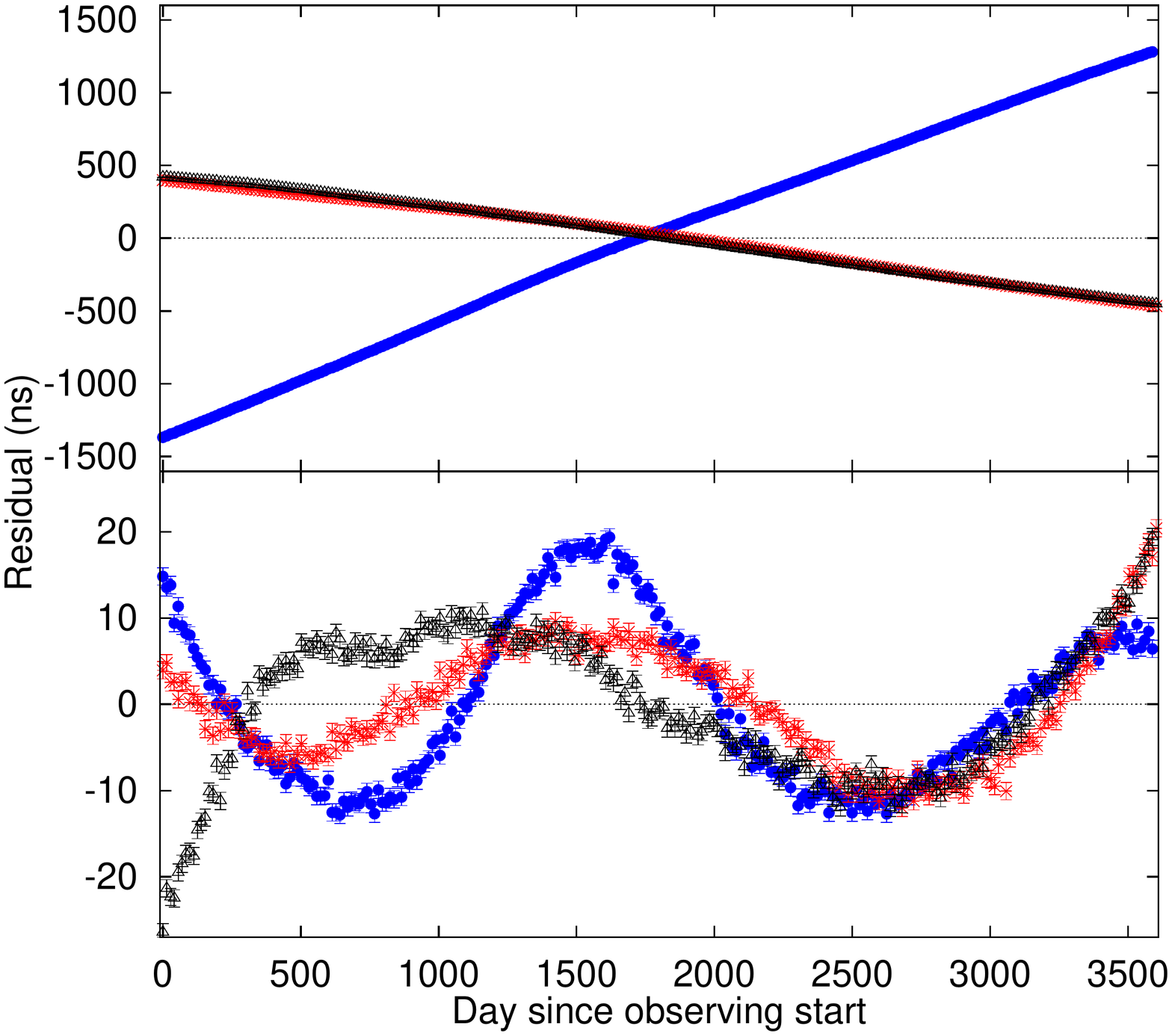}\\
\end{tabular}
\caption{Examples of timing residuals in three pulsars at different sky locations induced by selected GW signals plus noise. {\it Top left}: a continuous GW source generated by a circular SMBHB; {\it Top right}: a generic Gaussian burst; {\it Bottom left}: a burst with memory; {\it Bottom right}: a stochastic GWB from SMBHBs. In each panel top (bottom) plots show residuals before (after) fitting for the MSP spin and spindown. The fitting absorbs a large fraction of the signals with red spectra (from \protect\cite{2015arXiv151107869B}, courtesy of S. Burke-Spolaor).}
\label{fig_examplesignals}
\end{figure*}

The nominal frequency resolution of a PTA experiment is $\Delta{f}\approx1/T$, where $T$ is the duration of the experiment:
\begin{itemize}
\item If $(\omega_p-\omega)/(2\pi)>\Delta{f}$ for most MSPs, then there is no interference between the pulsar and the Earth terms; the latter can be added coherently and the former can be considered either as separate components of the signal or as an extra incoherent source of noise.
\item Conversely, if $(\omega_a-\omega)/(2\pi)<\Delta{f}$ for the majority of MSPs, then the pulsar terms add up to the respective Earth terms, affecting their phase coherency.  
\end{itemize}
This distinction has an impact on the detection strategy; different techniques are better suited to either situation, and many different detection algorithms have been developed accordingly, as we will see in Section \ref{sec:analysis}. Examples of timing residuals from a circular SMBHB are shown in the upper left panel of figure \ref{fig_examplesignals}; note that the signals are not perfect sinusoids because of the effect of the lower frequency pulsar term.

\vspace{0.3cm}
\hspace{-0.4cm}{\bf II - Generic bursts}\\
Bursts are generally defined as signals that are well localized in time, i.e., lasting much shorter than the observation time $T$. Note that PTAs are sensitive to nHz-$\mu$Hz frequencies, so that observable bursts will nevertheless last from weeks to several months. At such low frequencies and among the less exotic burst sources, we can expect defects to appear in a network of cosmic (super)strings when strings bend and reconnect, and which are known as cusps and kinks \cite{2001PhRvD..64f4008D}. For example, cusps have an extremely simple, linearly polarized waveform \cite{2003PhRvD..68h5017S,2009PhRvD..79d3014K}
\begin{equation}
h(t)=2\pi A|t-t_*|^{1/3}\,\,\,\,\,\,\,\,\,\,\,\,\, A\approx \frac{G\mu L^{2/3}}{D_c}
\end{equation}
where $G\mu$ is the string tension, $D_c$ is the (comoving) distance to the cusp and $L$ is its characteristic scale.

Another possible source of bursts consists of close encounters of SMBHs either on bound (elliptical) or unbound (parabolic, hyperbolic) orbits. Although the latter is extremely unlikely, the former might be a relatively common occurrence. It has, in fact, been shown that both three body scattering of ambient stars and torque exerted by a counter-rotating circumbinary disk can significantly increase the SMBHB eccentricity (see \cite{2012AdAst2012E...3D} and references therein). Another way to excite binary eccentricities is through the formation of a hierarchical SMBH triplet following two subsequent mergers \cite{2007MNRAS.377..957H,2016MNRAS.461.4419B}. Bursts of GWs can therefore be emitted by highly eccentric SMBHBs with orbital frequencies $\ll 1/T$ at periastron passage \cite{2010MNRAS.402.2308A}. Eccentricity causes a `split' of each polarization amplitude  $h_+(t)$ and $h_\times(t)$ into harmonics according to (see, e.g., equations (5-6) in \cite{2008PhRvL.100d1102W} and references therein):
\begin{eqnarray}
h^{+}_n(t) = A \Bigl\{-(1 + \cos^2\iota)u_n(e) \cos\left[\frac{n}{2}\,\Phi(t) + 2 \gamma(t)\right]\nonumber \\
\,\,\,\,\,\,\,\,-(1 + \cos^2\iota) v_n(e) \cos\left[\frac{n}{2}\,\Phi(t) - 2 \gamma(t)\right]
+ \sin^2\iota\, w_n(e) \cos\left[\frac{n}{2}\,\Phi(t)\right] \Bigr\},
\label{e:h+}\nonumber\\
h^{\times}_{n}(t) = 2 A \cos\iota \Bigl\{u_n(e) \sin\left[\frac{n}{2}\,\Phi(t) + 2 \gamma(t)\right] 
+ v_n(e) \sin\left[\frac{n}{2}\,\Phi(t) - 2 \gamma(t)\right] \Bigr\}\,.
\label{e:hx}
\end{eqnarray}
The coefficients $u_n(e)$,  $v_n(e)$, and $w_n(e)$ are linear combinations of Bessel functions of the first kind $J_{n}(ne)$, $J_{n\pm 1}(ne)$ and $J_{n\pm 2}(ne)$, and $\gamma(t)$ is an additional precession term in the phase given by $e$. For $e\ll 1$, $|u_n(e)| \gg |v_n(e)|\,,|w_n(e)|$ and the expressions above reduce to the circular case of equation (\ref{e:hpluscross1},\ref{e:hpluscross2}). Waveforms for parabolic SMBHB encounters are given in \cite{2010ApJ...718.1400F}. In general, a GW burst is detected as a short duration distortion in the timing residuals. In the upper right panel of figure \ref{fig_examplesignals}, we show an example of a burst with a generic Gaussian waveform.

\vspace{0.3cm}
\hspace{-0.4cm}{\bf III - Bursts with memory}\\
Besides the standard strain oscillation, GW bursts are also predicted to contain non-oscillatory components that result in a permanent deformation of spacetime. The final deformation depends on the radiation history of the source, and is therefore referred to as memory \cite{1991PhRvL..67.1486C,1992PhRvD..46.4304B,2009PhRvD..80b4002F}. Bursts displaying these features are known as bursts with memory (BWM). When the bursting source is a gravitationally-bound system, the memory arises from the fact that the radiated GW energy causes permanently non-vanishing second time derivatives in the source mass-energy quadrupole moments. Because of this, the spacetime metric relaxes to a configuration that differs from the pre-burst one. The merger of SMBHBs provides the most promising source of BWM for PTAs. The permanent displacement in the spacetime metric for SMBHBs inspiralling in a quasi-circular orbit up to the merger has a vanishing $h_\times$ component and takes the approximate form \cite{2014ApJ...788..141M}
\begin{equation}
  h_{+}\approx \frac{1-\sqrt{8}/3}{24}\frac{\mu}{D_l}\sin^2\iota(17+\cos^2\iota)\left[1+{\cal O}(\mu^2/M^2)\right]\approx 1.5\times 10^{-15}\frac{\mu}{10^9\msun}\left(\frac{D_l}{1{\rm Gpc}}\right)^{-1},
  \label{hmem}
\end{equation}
where $\mu$ is the redshifted reduced mass, $\iota$ is the inclination angle just prior to the final merger, and in the last passage, we have averaged over source inclinations. This displacement quickly arises as a few \% of the binary mass is radiated into GWs in the very last phase of the merger. The characteristic growth timescale is $t_r\approx 2\pi R_S/c\approx 1{\rm day}\,M_9$, where $M_9$ is the mass of the merger remnant in units of $10^9\msun$ and $R_s$ its Schwarzschild radius. After quickly ramping up, the perturbation simply settles to a constant value $h$. So for any practical purpose, the waveform is described by $h(t)=h_{+}\Theta(t-t_0)$, where $\Theta(t-t_0)$ is the heaviside step function. The perturbation is therefore null until time $t_0$, and quickly jumps to the value given by equation (\ref{hmem}), when the wave generated at the merger propagates through the detector. We can use equation (\ref{e:z1}) to get the associated redshift $z(t,\hat\Omega)$ and integrate over time according to equation (\ref{res}) to get the residual in the form
\begin{equation}
 r(t)=\frac{1}{2}\cos(2\psi)(1-\cos\theta)h_+[(t-t_e)\Theta(t-t_e)-(t-t_p)\Theta(t-t_p)].
\end{equation}
and $t_e-t_p\equiv\tau$ defined above. Since it is the integral of a constant, the residuals simply show a linear increase. Note that in this case we have a pulsar term, which is triggered at the time $t_p$ at which the wave `hits the pulsar', and an Earth term, which is triggered at the time $t_e$ at which the wave `hits the Earth'. As in the SMBHB case seen before, in a PTA, the Earth term will be correlated among all pulsars in the array, while the pulsar term will not. Contrary to the monochromatic waves, however, the two terms in general {\it do not} contribute to the detected signal at the same time. This is because $t_e-t_p$ is typically $\gg T$, the duration of the PTA experiment (unless the source is almost aligned with the considered pulsar). To imprint a signature onto the detected residuals, the `trigger' time must occur within the duration $T$ of the experiment. If this is not the case, then the signature is a continuous linear drift which is inevitably absorbed in a small correction to the pulsar frequency $\nu_0$. Examples of BWM are shown in the lower left panel of figure \ref{fig_examplesignals}. Contrary to the continuous wave case, the burst effect is largely absorbed by fitting for pulsar spin and spin derivative, which subtracts a quadratic function from the TOAs. 

\subsubsection{Stochastic backgrounds}
Stochastic GWBs in the PTA band can arise from a number of cosmological and astrophysical sources. As a first approximation, many calculations predict a characteristic strain spectrum with a single power-law shape
\begin{equation}
h_c=A\left(\frac{f}{\mathrm{yr}^{-1}}\right)^{-\alpha},
  \label{hcA}
\end{equation}
where $A$ is the strain amplitude at a reference frequency of 1yr$^{-1}$. The slope $\alpha$ differs depending on the specific background. On the cosmological side, cosmic string networks generate spectra with $\alpha=-5/3, -7/6, -1$ depending on several parameters defining the nature of the network \cite{2001PhRvD..64f4008D,2010PhRvD..81j4028O}. Standard inflation predicts $\alpha=-1$ with a signal amplitude that is well below foreseeable detection possibilities, even though several mechanisms can enhance the signal to detectable levels (see reviews in \cite{2016arXiv160501615C,2016JCAP...12..026B}). Other inflationary relics can produce stronger GWBs, with $0.5<\alpha<2$ \cite{2005PhyU...48.1235G}. Further cosmological GWBs include primordial BHs \cite{2011PhRvD..83h3521B} or QCD phase transitions, and may have more complicated spectra \cite{2010PhRvD..82f3511C}. The most promising signal for PTAs is, however, of astrophysical origin and stems from the cosmic population of SMBHBs \cite{1995ApJ...446..543R,2003ApJ...583..616J,2004ApJ...611..623S}. 

\begin{figure*}
\centering
\begin{tabular}{cc}
  \includegraphics[width=7.0cm,clip=true,angle=0]{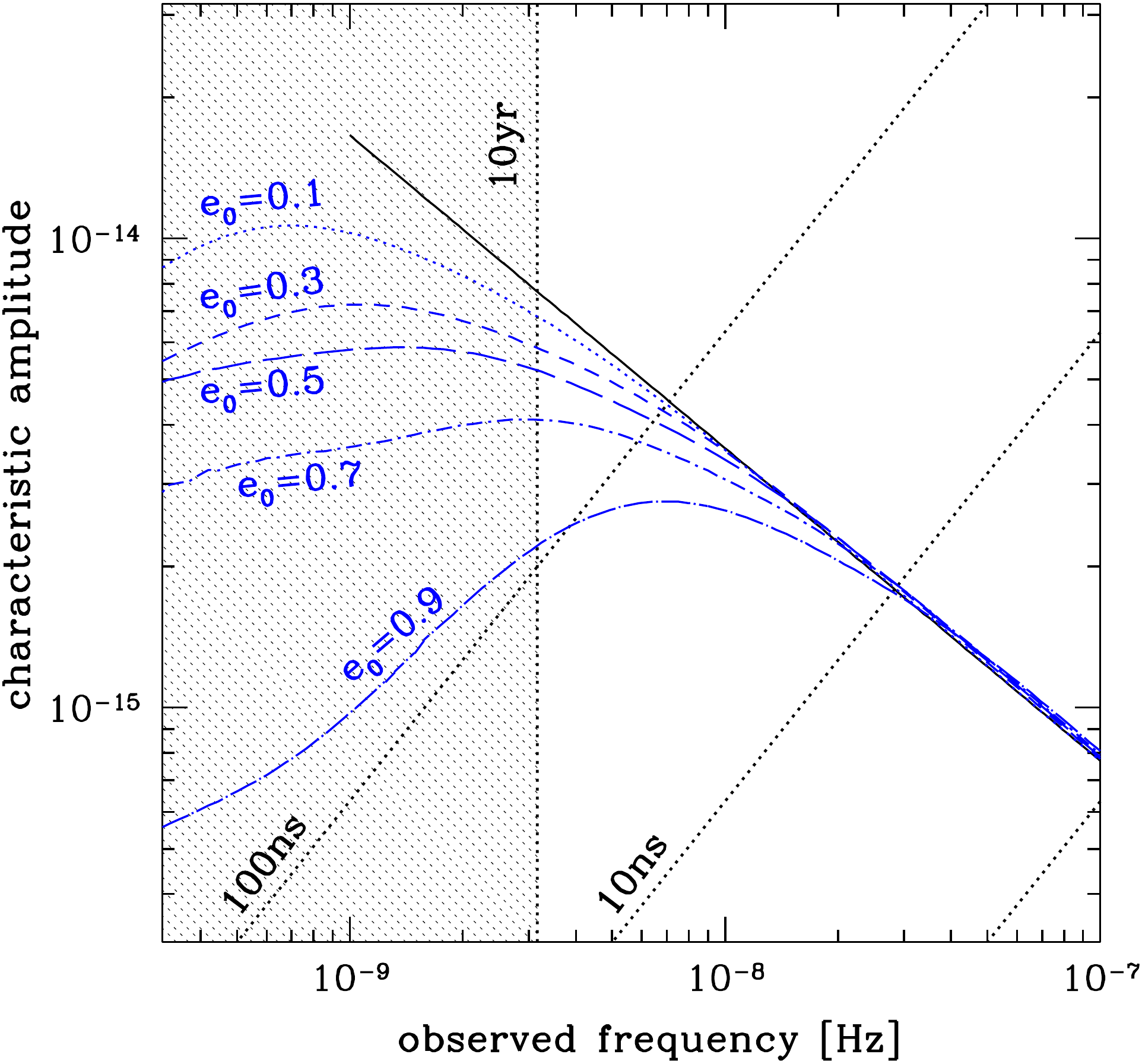}&
  \includegraphics[width=7.0cm,clip=true,angle=0]{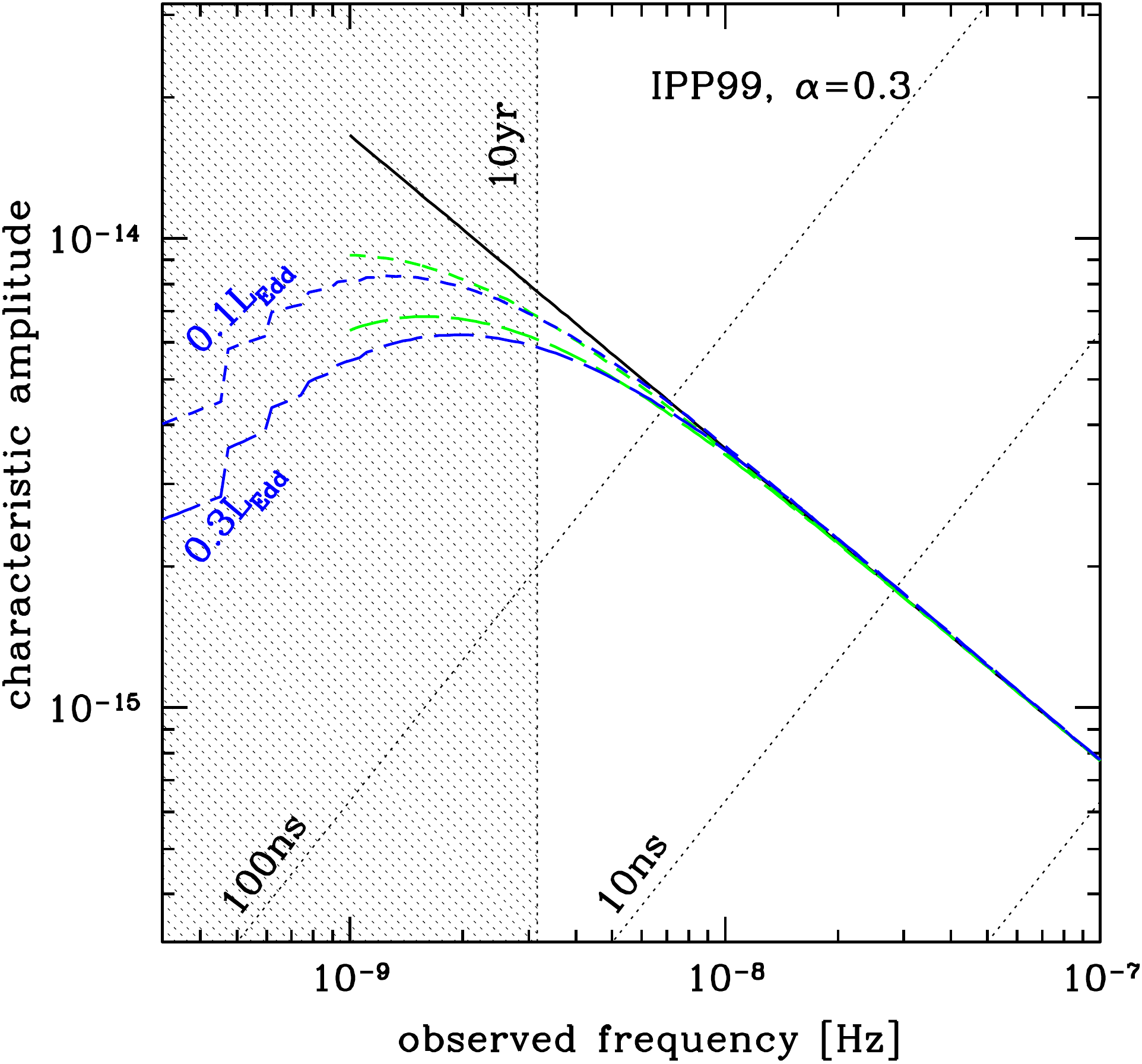}\\
\end{tabular}
\caption{Effect of the environment on $h_c$. {\it Left panel}: SMBHBs driven by stellar scattering; each blue line represents a population with a specific initial eccentricity, according to the model presented in \protect\cite{2010ApJ...719..851S}; {\it Right panel}: SMBHBs driven by interaction with a circumbinary disk modelled as in \protect\cite{1999MNRAS.307...79I}. Green lines are for circular binaries, blue lines allow for a self-consistent eccentricity evolution (models from \protect\cite{2011MNRAS.411.1467K}). The upper and lower pairs of curves are for two different Eddington ratios as labelled in the figure. In both panels, the solid black lines represent the GW driven case $h_c\propto f^{-2/3}$, dashed lines mark residual levels according to $r=h/(2\pi f)$ to guide the eye, and the excluded region at $f<3\times10^{-9}$ highlights the signal modification relevant to a PTA observation of $T \sim 10$yr. From \protect\cite{2015ASSP...40..147S}.}
\label{fig_stargas}
\end{figure*}

Since galaxy mergers are common \cite{1993MNRAS.262..627L}, we expect a large population of SMBHB to emit GWs in the PTA band {\it at any time} \cite{2008MNRAS.390..192S}. The superposition of many incoherent signals results in a GWB that is described by equation (\ref{hcA}), with $\alpha=-2/3$ \cite{2001astro.ph..8028P}. The normalization $A$ is affected by the poorly known SMBHB cosmic merger rate, but is predicted to be in the range of $10^{-16}<A<$few$\times 10^{-15}$ \cite{2013MNRAS.433L...1S,2014ApJ...789..156M,2015ApJ...799..178K,2015MNRAS.447.2772R,2016MNRAS.463L...6S}. Following \cite{2008MNRAS.390..192S} and assuming circular binaries, this stochastic GWB can be written in the form
\begin{equation}
h_c^2(f) =\int_0^{\infty} 
dz\int_0^{\infty}d{\cal M}\, \frac{d^3N}{dzd{\cal M} d{\rm ln}f}\,
h^2(f).
\label{hch2}
\end{equation}
where $h$ is the sky and polarisation averaged strain amplitude given by \cite{1987thyg.book..330T}
\begin{equation}
h={8\pi^{2/3}\over 10^{1/2}}{{\cal M}^{5/3}\over D_L}f^{2/3}\,,
\label{eqthorne}
\end{equation}
and the number of SMBHBs emitting per unit mass, redshift and log frequency is given by
\begin{equation}
\frac{d^3N}{dzd{\cal M}d{\rm ln}f} = \frac{d^2n}{dzd{\cal M}}\frac{dt_{\rm rf}}{d{\rm ln}f_{\rm rf}}
\frac{dz}{dt_{\rm rf}}\frac{dV_c}{dz}\,.
\label{d3N}
\end{equation}
In equation (\ref{d3N}), ${d^2n}/{dzd{\cal M}}$ is the cosmic merger rate density of SMBHBs, ${dt_{\rm rf}}/{d{\rm ln}f_{\rm rf}}$ is the time each binary spends in a given log frequency bin, and the other terms are standard cosmological relations. The level of the stochastic GWB therefore depends on the cosmic merger rate {\it and} on the mechanism driving the binary evolution through the ${dt_{\rm rf}}/{d{\rm ln}f_{\rm rf}}$ term. For GW driven binaries, ${dt_{\rm rf}}/{d{\rm ln}f_{\rm rf}}\propto f^{-8/3}$ and one recovers the $h_c\propto f^{-2/3}$ spectrum. However, since the GW emission efficiency has a steep $f$ dependence, at the relatively large separations relevant to PTA (centi-parsec), binaries might be still driven by the interaction with their stellar and gaseous environments \cite{2011MNRAS.411.1467K,2014MNRAS.442...56R,2017MNRAS.464.3131K}. For typical astrophysical systems (see derivation in \cite{2013CQGra..30v4014S}), the transition frequency between gas/star and GW dominated evolution is: 
\begin{eqnarray}
f_{{\rm star/GW}}\approx 5\times10^{-9}M_8^{-7/10}q^{-3/10} {\rm Hz}\nonumber\\
f_{{\rm gas/GW}}\approx 5\times10^{-9}M_8^{-37/49}q^{-69/98} {\rm Hz},
\label{decoup}
\end{eqnarray}
which is potentially within the PTA range. If binaries are eccentric, things are further complicated by the fact that each system emits a series of harmonics. A full mathematical derivation including stellar coupling and eccentricity can be found in \cite{2016arXiv160607484R}. The general effect of coupling with the environment is thus to produce a turnover of the spectrum at low frequencies, as shown in figure \ref{fig_stargas} for selected models. Future detailed measurement of the GWB spectral shape and normalization with PTAs can therefore in principle constrain the cosmic merger rate of SMBHBs, their dynamical interaction with the environment and their eccentricity distribution \cite{2017MNRAS.468..404C}.

\subsection{PTA sensitivity to gravitational waves}
The major challenge of PTAs is to dig out a possible GW signal (whether deterministic or stochastic) from a plethora of noise sources, many of which are poorly understood. The output of a detector can in fact be written as
\begin{equation}
  d(t)=s(t)+n(t),
\end{equation}
where $d(t)$ is the recorded data, $s(t)$ is the putative GW signal and $n(t)$ is the detector noise. In Fourier space, for a Gaussian stationary noise, $n(t)$ satisfies the ensemble average condition \cite{2000PhR...331..283M}
\begin{equation}
  \langle \tilde{n}^*(f)\tilde{n}(f')\rangle=\delta(f-f')\frac{1}{2}P_n(f)\,,\,\,\,\,\,\,\,\,\langle n^2(t)\rangle=\int_0^\infty df P_n(f),
\end{equation}
where $P_n(f)$ is the noise spectral density\footnote{Note that the spectral density is usually referred to as $S(f)$. In our notation, $S(f)[{\rm s}]$ is used in relation to the dimensionless GW strain. PTAs, however, measure TOAs and the associated power spectral density, denoted here as $P(f)$, has dimension of $[{\rm s}^{3}]$. The relation between the two is $P(f)=S(f)/(12\pi^2 f^2)$.}, which has been defined for positive frequencies $0<f<\infty$. In practice, the detectability of a signal depends on the noise spectral density $P_n(f)$ and on how it compares with the GW signal.

\subsubsection{Sources of noise in pulsar timing arrays}
\label{sec:noise}
An excellent review of the main noise sources relevant to PTAs is given in \cite{2013CQGra..30v4002C}. In practice, we can write 
\begin{equation}
P_n(f)=P_{\rm wn}(f)+P_{\rm rn,ac}(f)+P_{\rm rn,c}(f),
\label{eqrho}
\end{equation}
where $P_{\rm wn}(f)$ is white noise, $P_{\rm rn,ac}(f)$ is achromatic red noise and $P_{\rm rn,c}(f)$ is chromatic red noise.

The power contributed by white noise takes the form
\begin{equation}
P_{\rm wn}(f)=2\sigma^2_{\rm wn}\Delta{t},
\label{eqrho}
\end{equation}
where $\Delta{t}$ is the cadence of observations (typically weeks) and $\sigma_{\rm wn}$ is the rms uncertainty in the TOA. The main sources of white noise in PTA observations are radiometer noise and jitter. Radiometer noise defines the maximum theoretical precision in measuring TOAs, and arises from the fact that folded pulses with finite S/N are matched to a theoretical template. Jitter is due to the intrinsic stochasticity of the phasing of individual pulses. Detail scaling for these noise sources is given in \cite{2013CQGra..30v4002C,2015JPhCS.610a2019W}; typical figures of $\sigma_{\rm wn}$ are hundreds of ns (radiometer) and tens of ns (jitter).

Chromatic red noise, by definition, depends on the frequency of the observed radio photons and arises from frequency-dependent propagation effects in the ISM, in particular dispersion and scattering. Interaction of radio photons with the ISM's cold magnetized plasma yields a frequency-dependent delay in their group velocity. This causes a delay in TOAs that is proportional to the electron column density (referred to as dispersion measure, DM) travelled by the radio photons with a $\nu_\gamma^{-2}$ dependence, where $\nu_\gamma$ is the frequency of the observed photons (not to be mistaken with the spinning frequency of the MSP, $\nu$). Scattering is the pulse broadening due to multiple paths travelled along the ISM and has a $\nu_\gamma^{-4}$ dependence. Note that, as both the Earth and the observed MSPs move in the Galaxy potential, the DM is typically time-dependent. Because of their frequency dependence, chromatic noise sources can be dealt with by using wideband receivers and fitting for the frequency dependence of the TOAs\footnote{Note that wideband observations entail other issues related to frequency dependence of the pulse profile. This can be dealt with, for example, by developing 2D (time-frequency) profile templates \cite{2014MNRAS.443.3752L,2014ApJ...790...93P}}. 

Conversely, achromatic red noise is the same at all received radio frequencies and cannot be mitigated by means of wideband observations. This noise is intrinsic to the pulsar and is due to the complex torques arising by crust-superfluid interactions. Spin noise has been detected in several MSPs and has a very steep red spectrum $P_{\rm sn}(f)\propto f^{-5\pm 0.4}$ \cite{2010ApJ...725.1607S}. For comparison, a GWB with $h_c\propto f^{-2/3}$ results in $P_h(f)\propto f^{-13/3}$ according to equation (\ref{e:Ph}). Spin noise can therefore be the most serious limiting factor for the detection of a stochastic GWB.  

The noise sources that were considered thus far are supposedly uncorrelated among MSPs. There are however additional sources of noise that show specific correlation patterns. Clock offsets have the same effect on all pulsars, and therefore induce a monopole correlated signal. Errors in the solar system ephemeris (which are necessary for computing TOAs at the solar system barycentre) result in a dipole correlation pattern \cite{2016MNRAS.455.4339T}. Fortunately, those are different from the quadrupole $\Gamma_{ab}$ correlation diagnostic for a stochastic GWB, and advanced analysis methods can distinguish between them \cite{2017PhRvD..95d2002T}.


\subsubsection{S/N calculation and scaling relations}
\label{sec:SNR}
With an understanding of the GW signature imprinted on timing residuals and of the relevant sources of noise, we can estimate typical signal-to-noise ratios (S/N, $\rho$) of different GW signals as a function of the structural array parameters and assess prospects for their detectability. In the following, we make the distinction between deterministic signals and stochastic GWBs.

\vspace{0.3cm}
\hspace{-0.4cm}{\bf I - Deterministic signals}\\
For deterministic signals, the data can be matched-filtered with a template for $s(t)$. It can be shown (e.g. \cite{2010PhRvD..81j4008S}) that in this case, the S/N of the GW signal is given by 
\begin{equation}
 \rho^2=(r(t)|r(t)) 
\end{equation}
where $(\cdot|\cdot)$ is the weighted inner product defined as
\begin{equation}
(x|y)  =  2 \int_{0}^{+\infty} \frac{\tilde x^*(f) \tilde y(f) +  \tilde x(f) \tilde y^*(f)}{P_n(f)} df\, \simeq \frac{2}{P_0}\int_0^{T} x(t) y(t) dt\,.
\label{e:innerxyapprox1}
\end{equation}
and
\begin{equation}
\tilde x(f) = \int_{-\infty}^{+\infty} x(t) e^{-2\pi i f t}.
\label{e:tildex}
\end{equation}
Note that in the last step of equation (\ref{e:innerxyapprox1}), we implicitly assumed that the signal is monochromatic, and $P_0$ is the noise spectral density at the frequency of the signal.
For an array with $M$ pulsars identified by index $a$, we have
\begin{equation}
\rho_a^2 = \frac{2}{P_{0,a}}\int_0^{T_a} r^2_a(t) dt,\,\,\,\,\,\,\,\,\rho^2=\sum_a \rho_a^2.
\label{e:innerxyapprox}
\end{equation}
The equation above also applies to a burst generated by very eccentric binaries by summing over all harmonics and considering the appropriate $P(f)$ at the observed frequency of each harmonic.

The integral in equation (\ref{e:innerxyapprox}) can be easily computed using the residual formula for a circular SMBHB given in equation (\ref{Eq:signal}). For simplicity, we only consider the Earth term, and an array of $M$ identical MSPs, dropping the $a$ index. Individually-resolvable sources are usually expected to be observed at $f\equiv\omega/(2\pi)\gg 1/T$; we therefore assume white noise, $P_{0}=2\sigma^2\Delta{t}$. Averaging over the antenna response functions $F^+, F^{\times}$ and orbital inclinations $\iota$, and summing over all MSPs, we get

\begin{equation}
\rho^2 \approx \frac{M}{15\pi^2}\frac{T}{\Delta t}\frac{A^2}{\sigma^2f^2}.
\label{rho2single}
\end{equation}
Noticing that $N=T/\Delta{t}$ is the number of observations and taking the square root we finally get
\begin{equation}
\rho \approx \frac{1}{\sqrt{15}\pi}\frac{A}{\sigma f}(NM)^{1/2}.
\label{rhosingle}
\end{equation}
The S/N of a circular SMBHB is therefore proportional to the square root of the number of pulsars in the array and of the number of observations (i.e. the total observation time $T$, for a uniform observation cadence), and is inversely proportional to the rms residual $\sigma$. Equation  (\ref{rhosingle}) can be inverted to obtain the minimum amplitude $A$ observable at a given S/N threshold:

\begin{equation}
A\approx {\sqrt{15}\pi}\rho {\sigma f}(NM)^{-1/2}=9\times 10^{-15}\,\frac{\rho}{5}\,\frac{\sigma}{100{\rm ns}}\,\frac{f}{10^{-7} {\rm Hz}}\left(\frac{N}{250}\right)^{-1/2}\left(\frac{M}{20}\right)^{-1/2}.
\label{aminCGW}
\end{equation}
Although SMBHBs are abundant in the Universe (see figure 1 in \cite{2012MNRAS.420..860S}), comparison between the above estimate and equation (\ref{e:Agw}) shows that current PTAs are only sensitive to extremely massive SMBHBs, which are extremely rare. Equation (\ref{aminCGW}) can be compared with the limits shown in figure \ref{fig_singleUL}. At $10^{-7}$Hz, the EPTA upper limit is $\approx 3\times 10^{-14}$, which is in line with the equation above, considering that the EPTA dataset is dominated by one pulsar with $\sigma=130$ns \cite{2016MNRAS.458.3341D}. We also note that the frequency dependence is somewhat flatter than $f$, indicating some red noise contribution. The turnover at $f<10^{-8}$ is instead due to a combination of red noise and MSP spin and spindown fitting.

A similar derivation for BWM can be found in \cite{2010MNRAS.401.2372V}, yielding
\begin{equation}
h_{\rm min}\approx {12\sqrt{3}}\rho {\sigma}T^{-1}(NM)^{-1/2}{\cal F}(\chi)=4.5\times 10^{-16}{\cal F}(\chi)\,\frac{\rho}{5}\,\frac{\sigma}{100{\rm ns}}\,\left(\frac{T}{10{\rm yr}}\right)^{-1}\left(\frac{N}{250}\right)^{-1/2}\left(\frac{M}{20}\right)^{-1/2},
\label{hminBWM}
\end{equation}
where ${\cal F}(\chi)$, given in \cite{2010MNRAS.401.2372V}, is a function of $\chi=t_0/T$ (being $t_0$ the BWM arrival time) and has a minimum value of $\approx 1.4$. When compared to equation (\ref{hmem}), the above estimate suggests that PTAs can be sensitive to BWM out to much larger distances than inspiralling SMBHBs. Note however that while inspiralling SMBHBs are rather abundant, coalescences are extremely rare events. In fact the coalescence rate of SMBHBs with $M>10^9\msun$ throughout the Universe is $<10^{-2}$yr \cite{2016MNRAS.456..961B}.

\vspace{0.3cm}
\hspace{-0.4cm}{\bf II - Stochastic backgrounds}\\
For stochastic signals, the strategy is to detect cross-correlated power in several detectors (i.e. in several pulsars). The S/N imprinted by a stochastic GWB in a PTA can be written as \cite{1999PhRvD..59j2001A,2000PhR...331..283M,2015MNRAS.451.2417R}
\begin{equation}
\rho^2=2\sum_{a=1,M}\sum_{b>a}T_{ab}\int \frac{\Gamma_{ab}^2P_h^2}{P_{n,ab}^2}df,
\label{eqrho}
\end{equation}
where the sums run over all pulsar pairs, $T_{ab}$ is the timespan for which both pulsars $a$ and $b$ are observed (note that, in general, MSPs have different time coverage, depending on when they were discovered, the schedule requirement at observatories, etc.) and $\Gamma_{ab}$ is the HD correlation function defined by equation (\ref{eq:HD}). The correlated noise term is given by
\begin{equation}
  P_{n,ab}^2=P_aP_b+P_h[P_a+P_b]+P_h^2(1+\Gamma_{ab})^2,
    \label{eqsn}
\end{equation}
where
\begin{equation}
\label{eq:pisimple}
P_{a,b}=2\sigma_{a,b}^2\Delta t+P_{{\rm rn},a,b},
\end{equation}
and $P_h$ is related to the GWB characteristic strain via equation (\ref{e:Ph}). In the following, we ignore the red noise component $P_{{\rm rn},a,b}$ for simplicity. Note that $P_{n,ab}$ has two distinct asymptotic trends. For $P_h\ll P_{a,b}$, i.e. in the weak signal limit, it reduces to $P^2_{n,ab}=P_aP_b$. However, when $P_h>P_{a,b}$, it does not matter how strong the signal is, the integrand of equation (\ref{eqrho}) is at most of the order $\Gamma_{ab}^2\ll 1$. This means that the PTA performance in terms of GWB detection depends on the strength of the signal, as pointed out in \cite{2013CQGra..30v4015S}.

\begin{figure*}
\centering
  \includegraphics[width=12.0cm,clip=true,angle=0]{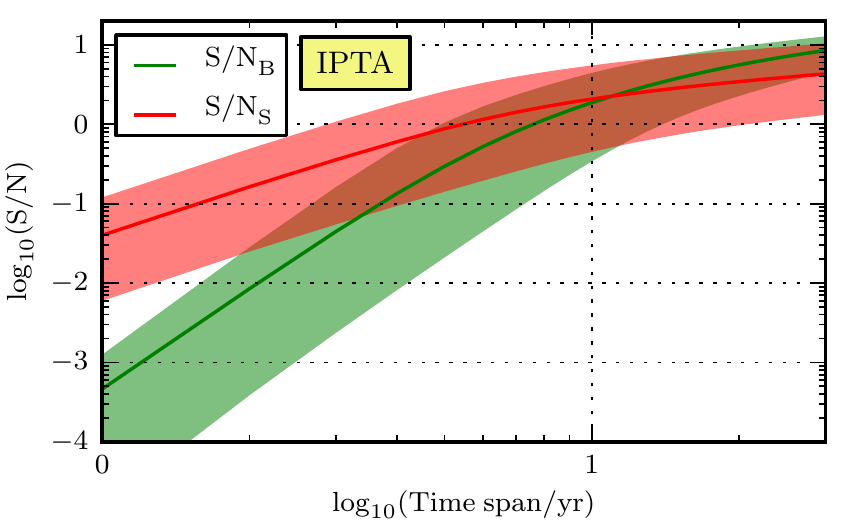}
\caption{S/N scaling with observation time of the loudest resolvable circular SMBHBs (red) and of the collective stochastic GWB (green) for $10^{5}$ realizations of the cosmic SMBHB population. The coloured band shows the 90\% confidence intervals, while the solid lines represent the median of all realizations. An IPTA-type array with 49 MSPs with $\sigma$ between 500ns and 9$\mu$s was assumed. Note the two distinct scalings of the GWB S/N in the weak and strong signal regimes. We also note that  the resolvable SMBHB S/N does not follow a simple $T^{1/2}$ scaling in this figure. This is because we do not plot the S/N of a specific source, but of the {\it loudest} source, which might change as lower frequencies (and better frequency resolution) are accessible with increasing $T$ (from \cite{2015MNRAS.451.2417R}).}
\label{fig_rhotime}
\end{figure*}

Let us again drop the $a,b$ indexes by considering equal MSPs in the array. We further assume $M\gg 1$, so that we can substitute $\Gamma_{ab}$ with the average value $\Gamma=1/(4\sqrt{3})$ to get 
\begin{equation}
\rho^2\approx T\Gamma^2M^2\int \frac{P_h^2}{P_n^2} df.
\label{eqrhosimpl}
\end{equation}
In the limit $P_h\ll P_{n}$ and writing $h_c$ according to equation (\ref{hcA}), after some algebra and integrating in the range $1/T<f<\infty$ we get
\begin{equation}
\rho \approx 2\times 10^{-13}NM\sigma^{-2}A^2T^{10/3}.
\label{eqrhoweak}
\end{equation}
The array sensitivity therefore increases quickly by improving timing precision and by extending the time baseline of the experiment (see figure \ref{fig_rhotime}). Observing more pulsars also help, but only linearly in S/N. The minimum detectable GWB therefore has a characteristic strain $A$ given by
\begin{equation}
A\approx 7\times 10^6 \rho^{1/2}\sigma T^{-5/3} (NM)^{-1/2} \approx 10^{-16} \,\rho^{1/2}\,\frac{\sigma}{100{\rm ns}}\,\left(\frac{T}{10{\rm yr}}\right)^{-5/3}\left(\frac{N}{250}\right)^{-1/2}\left(\frac{M}{20}\right)^{-1/2}.
\label{eqaweak}
\end{equation}
Note that although this lies at the lower end of the strain range predicted by cosmological models of SMBH assembly \cite{2013MNRAS.433L...1S,2014ApJ...789..156M,2015ApJ...799..178K,2015MNRAS.447.2772R,2016MNRAS.463L...6S}, there are strong caveats. First, we ignored both red noise and MSP spin and spin-down fitting. The latter generally compromises the PTA sensitivity at $f<2/T$ (see e.g. figure 1 in \cite{2013CQGra..30v4002C}) thus degrading the above estimate by a factor of a few. Second, when $\rho=1$ we are already departing from the weak signal regime. 

When $P_h> P_{n}$, things are drastically different. The integral in equation (\ref{eqrhosimpl}) reduces to $\int df$ giving
\begin{equation}
\rho \approx T^{1/2}\Gamma M(f_{\rm max}-f_{\rm min})^{1/2}\approx \Gamma M \Sigma^{1/2}.
\label{eqrhoweak}
\end{equation}
Here $f_{\rm min}=1/T$ and $f_{\rm max}$ is the maximum frequency for which the condition $P_h> P_{n}$ is satisfied. In the last step, we divided the frequency domain in resolution bins $\Delta{f}=1/T$, and $\Sigma$ is the number of frequency bins for which $P_h> P_{n}$ (usually a few). We now see that the S/N increases linearly with the number of pulsars in the array $M$  and, as shown in figure \ref{fig_rhotime}, only with the square root of the observation time. Timing precision only plays a minor role by weakly affecting the $f_{\rm max}$ limit (or alternatively $\Sigma$). In practice, to make a confident detection of a stochastic GWB, it is absolutely necessary to include a large number of quality MSPs in the array. Note that, although we assumed $h_c\propto f^{-2/3}$, the derived scalings usually hold for any GWB shape, as shown in \cite{2016PhRvD..94l3003V} for broken power-law spectra approximating the GW signals from SMBHBs interacting with gas or stars.

\subsection{Current status of gravitational wave searches}
\label{subsec:3.2}

\subsubsection{Analysis methods}
\label{sec:analysis}
Whether the signal is deterministic or stochastic, the challenge of data analysis is to determine what the chances are that it is present in the data. The problem can be tackled either using a frequentist or a Bayesian philosophy. Reviewing principles and differences of those approaches is well beyond the scope of this contribution; we summarize here the main ideas and point the reader to the relevant PTA literature. The core aspect of all modern GW searches is the evaluation of the likelihood  that some signal is present in the time series of the pulsar TOAs. Deferring technical details to e.g. \cite{2013MNRAS.428.1147V,2015MNRAS.453.2576L}, the likelihood marginalised over the timing parameters 
can be written as

\begin{eqnarray}
        \mathcal{L}(\vec{\theta}, \vec{\lambda}, \vec{\eta} | \vec{\delta t}) 
        = \frac{1}{\sqrt{(2\pi)^{n-m} det(G^T C(\vec{\eta},\vec{\theta}) G) }} \times
\nonumber \\
\exp{\left(-\frac1{2} (\vec{\delta t} - \vec{r}(\vec{\lambda}))^T G 
(G^T C(\vec{\eta},\vec{\theta}) G)^{-1} G^T (\vec{\delta t} - \vec{r}(\vec{\lambda})) \right)}.
\label{Eq:lik}
\end{eqnarray}

Here $n$ is the length of the vector $\vec{\delta t} = \cup \delta{t}_{a}$ obtained by concatenating the individual pulsar TOA series $\delta{t}_{a}$, $m$ is the number of parameters in the timing model, and the matrix $G$ is related to the design matrix (see \cite{2013MNRAS.428.1147V} for details). $\vec{\theta}, \vec{\lambda}, \vec{\eta}$ are vectors of parameters describing the noise ($\vec{\theta}$), a deterministic GW signal ($\vec{\lambda}$) or the spectral shape of a stochastic GWB ($\vec{\eta}$). In general, the variance-covariance matrix $C$ contains contributions from the putative GWB and from white and red noise: $C = C_{gw}(\vec{\eta}) + C_{wn}(\vec{\theta}) + C_{rn}(\vec{\theta})$. The detailed form of all the contributions to the variance-covariance matrix can be found in \cite{2015MNRAS.453.2576L}. 

In frequentist searches, a detection statistic is constructed based on the likelihood function both in the null hypothesis and in the presence of a signal. Extensive Monte-Carlo simulations on synthetic data with injected signals are then performed to construct the detection probability as a function of the false alarm rate. Comparison with the value of the statistic obtained from the real dataset is then used to either claim a detection (with associated confidence) or to obtain an upper limit in case of no detection. This procedure has been detailed in \cite{2012ApJ...756..175E} and has been used in several deterministic source searches (e.g. \cite{2014MNRAS.444.3709Z,2016MNRAS.455.1665B}).

In Bayesian searches, the likelihood function is used to compute the odds ratio of the Bayesian evidence for the hypothesis that a signal is present in the data (model $\mathcal{H}_1$, with signal described by parameters $\vec{\lambda}$ and/or $\vec{\eta}$) versus the null hypothesis ($\mathcal{H}_0$). In the case of no prior preference of either model, the odds ratio reduces to the Bayes factor, $\mathcal{B}$:   
\begin{equation}
\mathcal{B}= \frac{\int \mathcal{L}( \vec{\theta}, \vec{\lambda}| \vec{\delta t} )\pi(\vec{\theta}, \vec{\lambda})d\vec{\theta}\,d\vec{\lambda}}{\int \mathcal{L}( \vec{\theta}| \vec{\delta t})\pi(\vec{\theta})d\vec{\theta}},
\end{equation}
which is technically the ratio of the evidences for the hypothesis $\mathcal{H}_1$ and $\mathcal{H}_0$. The value of $\mathcal{B}$ is the statistic used to assess the presence of the signal \cite{kassr95}. In the case of a detection, the shape of the likelihood function can be used to infer the parameters of the signal and their uncertainties, otherwise upper limits can be placed. Bayesian searches have been recently extensively applied to the search for both deterministic signals and stochastic GWBs. 

\subsubsection{Overview of current results}
The search for deterministic signals in real data have so far focused on circular SMBHBs and BWM. In modern algorithms, a deterministic signal $r(\vec{\lambda},t)$ is added to the model and a search is performed over the parameter space defined by $\vec{\lambda}$. For individual SMBHBs, $\vec{\lambda}$ includes the source amplitude, frequency, sky location, inclination phase and polarization (plus other parameters related to the pulsar term, when included in the search); for BWM, parameters include burst amplitude and trigger time $t_0$, sky location, and source inclination.

\begin{figure}
\centerline{\includegraphics[width=13cm,angle=0,clip=1]{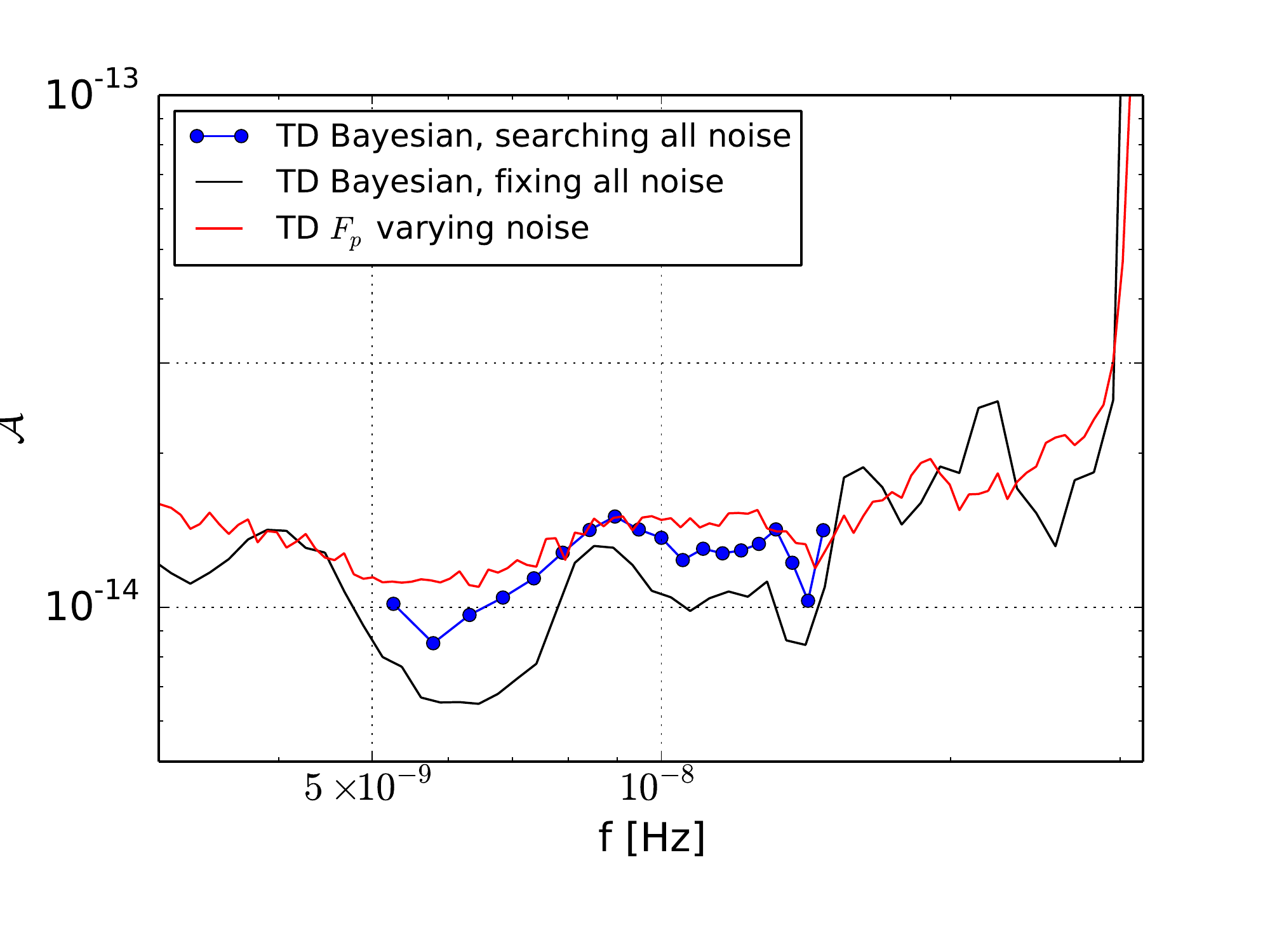}}
\caption{Sky averaged 95\% upper limit on the gravitational wave amplitude $A$ of a circular SMBHB as a function of frequency, placed by three different searches (labelled in figure) performed on the EPTA dataset (from \cite{2016MNRAS.455.1665B}, where the detailed descriptions of each method can be found).}
\label{fig_singleUL}
\end{figure}

\cite{2010MNRAS.407..669Y} obtained the first frequentist individual SMBHB upper limit on early PPTA data by looking for excess power as a function of frequency, placing a sky averaged upper limit on the source amplitude (cf equation \ref{e:Agw}) of $A\approx 10^{-13}$ at 10 nHz. More recently, detection statistics for single SMBHBs have been calculated for circular systems either including the Earth term only \cite{2012PhRvD..85d4034B}, or adding the pulsar term \cite{2012ApJ...756..175E}, as well as for eccentric binaries \cite{2016ApJ...817...70T}. Alternative frequentist methods based on the construction of null streams have also been proposed \cite{2015MNRAS.449.1650Z}. In parallel, \cite{2014PhRvD..90j4028T} developed a Bayesian pipeline that can handle generic circular SMBHBs. Searches on real data have been performed by the three major PTAs \cite{2014ApJ...794..141A,2014MNRAS.444.3709Z,2016MNRAS.455.1665B}, yielding null results. The EPTA placed the most stringent limit to date, shown in figure \ref{fig_singleUL} (from \cite{2016MNRAS.455.1665B}). Around 10 nHz, sources with $A> 10^{-14}$ can be confidently excluded. Compared to equation (\ref{e:Agw}) this rules out the presence of centi-parsec SMBHBs of a few billion solar masses out to the distance of the Coma cluster. Note that those limits are consistent with our current understanding of SMBH assembly, as state-of-the-art models predict a mere 1\% chance of making a detection at this sensitivity level \cite{2016MNRAS.455.1665B}.

Searches for BWM have been performed both with the PPTA \cite{2015MNRAS.446.1657W} and NANOGrav \cite{2015ApJ...810..150A} datasets. Both searches yielded comparable results, constraining the BWM rate to be less than $\approx 1{\rm yr}^{-1}$ at $h=10^{-13}$. To produce a strain of comparable amplitude, a $10^9\msun$ SMBHB should merge in the Virgo cluster, which is an extremely unlikely event. In fact, \cite{2012ApJ...752...54C,2015MNRAS.447.2772R} estimated that the event rate for such a strong burst is $<10^{-6}{\rm yr}^{-1}$, which makes these null results unsurprising.

\begin{figure}
\centerline{\includegraphics[width=13cm]{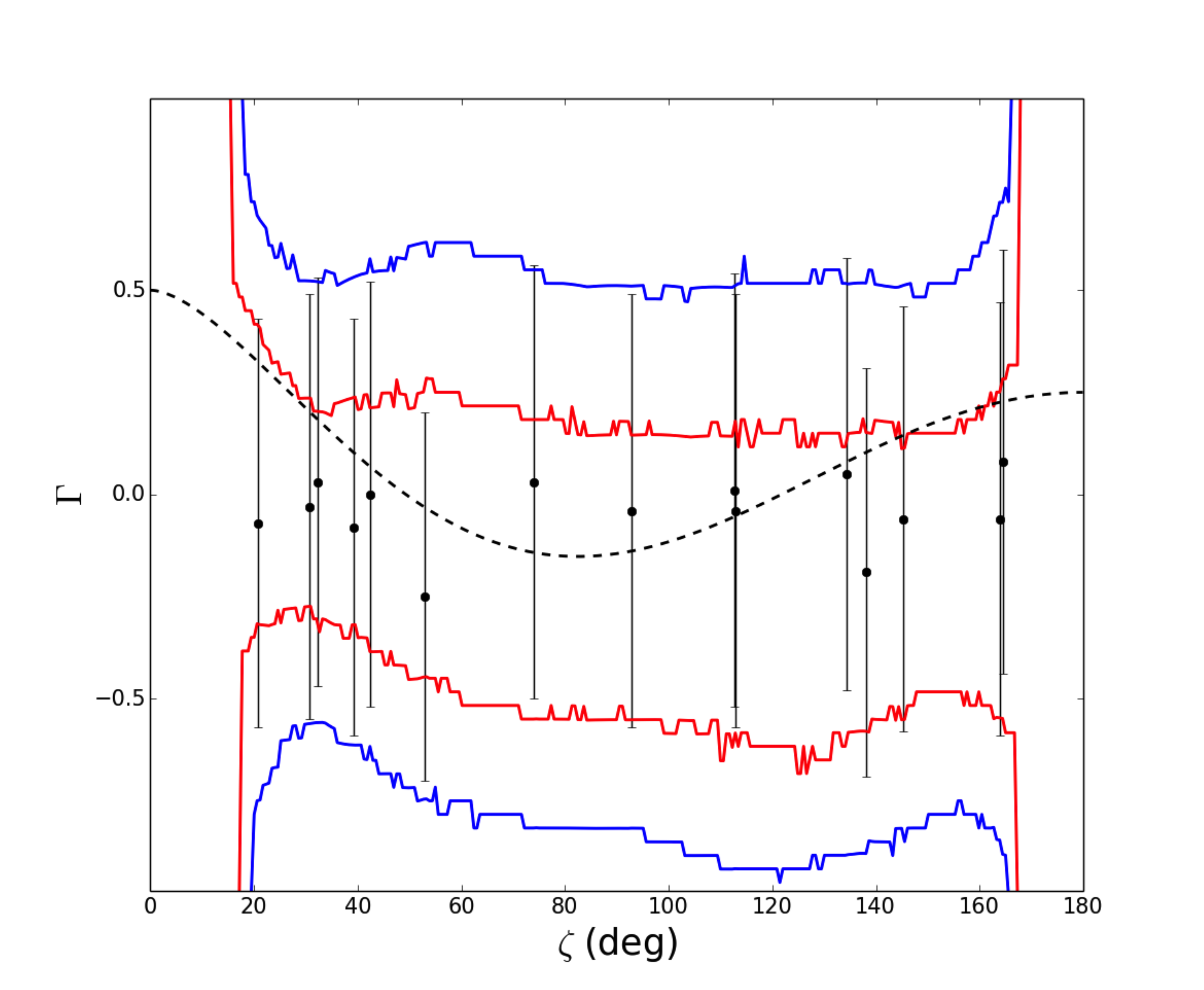}}
\caption{The recovered correlation between pulsars as a function of angular separation on the sky in the EPTA analysis. The red and blue lines represent the $68\%$ and $95\%$ confidence intervals of the recovered correlation. Individual points represent the mean correlation coefficients with a $1\sigma$ uncertainty for each pulsar pair. The dashed line represents the HD correlation (from \cite{2015MNRAS.453.2576L}).}
\label{fig_correlation}
\end{figure}

\begin{figure}
\centerline{\includegraphics[width=11.8cm,angle=0,clip=0]{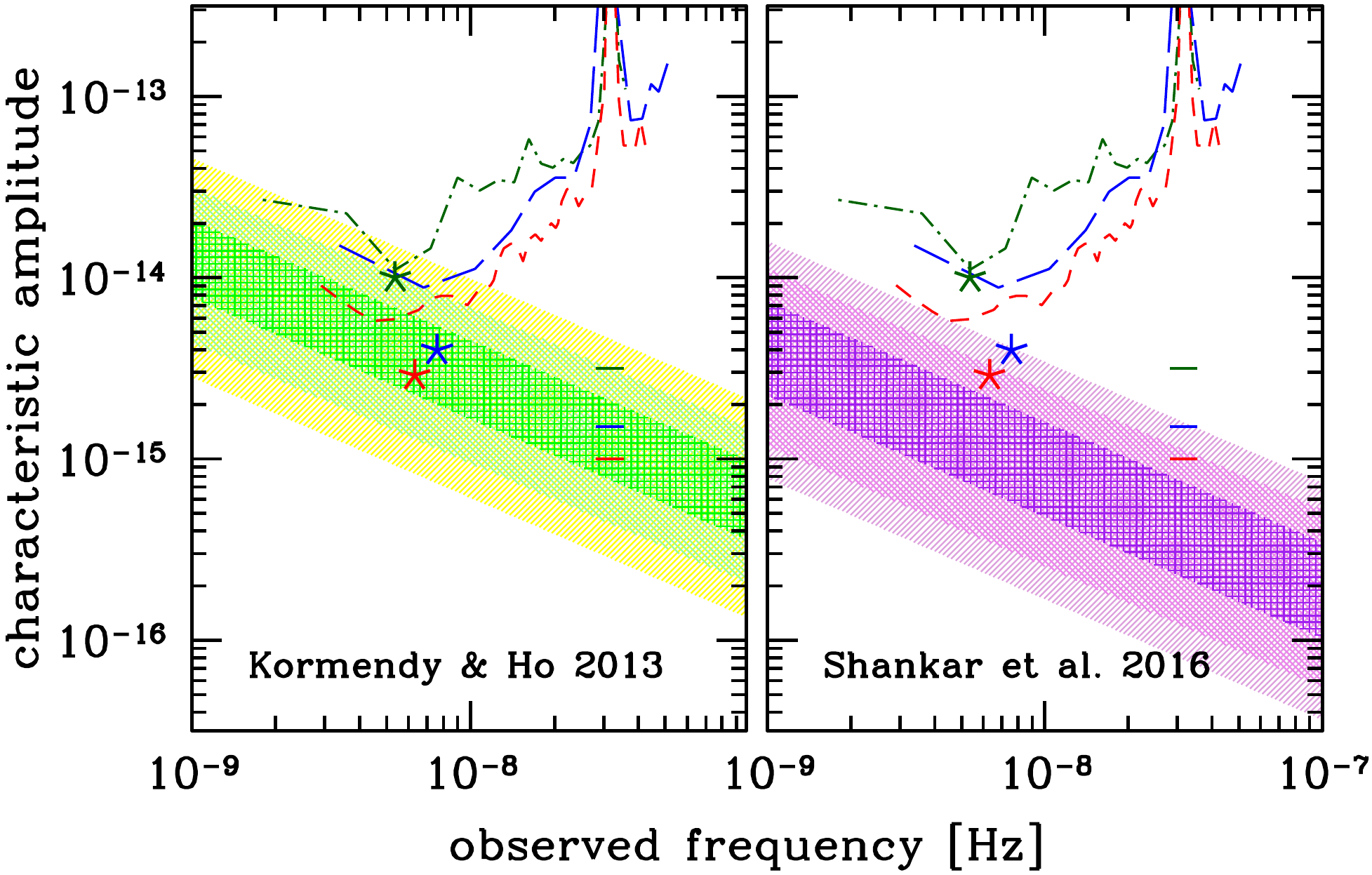}}
\caption{ Current upper limits on stochastic GWBs as a function of frequency for EPTA (dot-dashed green), NANOGrav (long-dashed blue), and PPTA (short-dashed red). For each curve, stars represent the integrated limits to an $f^{-2/3}$ background, and horizontal ticks represent their extrapolation at $f=1$yr$^{-1}$, i.e. the upper limit on $A$ defined by equation (\ref{hcA}). Shaded areas represent the 68\% 95\% and 99.7\% confidence intervals of $h_c(f)$ for selected SMBHB population models. The two panels show how uncertainties in the SMBH mass-host galaxy relation severely impact the expected signal level. See \cite{2016MNRAS.463L...6S} for full details of the employed models (from \cite{2016MNRAS.463L...6S}).}
\label{fig_hclimit}
\end{figure}

All PTAs (including the IPTA) performed extensive searches for a stochastic GWB from SMBHBs, which is the most likely GW signal to be first detected by PTAs \cite{2015MNRAS.451.2417R}. In isotropic GWB searches, the smoking gun of a detection is provided by the HD correlation pattern given by equation (\ref{eq:HD}). The correlation can be used to construct an optimal correlation statistics based on the maximum likelihood estimator \cite{2009PhRvD..79h4030A,2015PhRvD..91d4048C}, which can be employed to obtain frequentist upper limits \cite{2015MNRAS.453.2576L}. In advanced algorithms, the HD correlation is included in the analysis via the correlation matrix $C_{gw}(\vec{\eta})$, which, in the often used Fourier representation, takes the form (e.g. \cite{2015MNRAS.453.2576L})

\begin{eqnarray}
\label{Eq:FreqMatrix}
\Psi_{a,b,i,j} = \Gamma(\zeta_{ab}) \varphi^{\mathbf{\mathrm{GWB}}}_{i}\delta_{ij},
\end{eqnarray}
where indices $i,j$ run over the difference frequency bins of the Fourier decomposition and $\varphi_i$ is the power in the GW signal given by equation (\ref{e:Ph}), which is evaluated at the central frequency $f_i$ if the $i$-th bin. The matrix (\ref{Eq:FreqMatrix}) is included in the appropriate Fourier representation of the likelihood function (\ref{Eq:lik}). Note that in the case of {\it anisotropic} GWB, the signal can be decomposed into spherical harmonics, and the power in different harmonics has different correlation patterns (since the HD curve is the correlation of the monopole component) that can also be included in the analysis \cite{2014PhRvD..90h2001G,2015PhRvL.115d1101T}. 

In the simplest searches, the $h_c(f)$ responsible for $P_h(f)$ is described by a single power law defined by the two parameters $\vec{\eta}=A,\alpha$ (cf equation (\ref{hcA})), however additional parameters can be (and have been) included in the search to describe a low frequency turnover. In the following, we always refer to the upper limit placed on $A$ assuming an $f^{-2/3}$ GWB, appropriate for circular, GW-driven SMBHBs. Note that those can be easily converted into limits on $\Omega_{\rm gw}$ by combining equations (\ref{eq:omegagw}) and (\ref{hcA}). Systematic searches for GWBs in PTA data have been ongoing for more than a decade \cite{2006ApJ...653.1571J,2011MNRAS.414.3117V,2013ApJ...762...94D}, with early upper limits on $A$ in the range $6\times10^{-15}-1.1\times 10^{-14}$, which is still higher than the range predicted by theoretical models. In recent years, improvements in the data quality and in the search algorithms allowed us to push the sensitivity down to $A\approx 10^{-15}$, yielding null results. Using the first EPTA legacy data release, \cite{2015MNRAS.453.2576L} placed an upper limit of $A=3\times10^{-15}$ for a SMBHB GWB, while also providing upper limits on the cosmic string tension of $G\mu=1.1\times 10^{-7}$, which is competitive with CMB constraints \cite{2016A&A...594A..13P}. Figure \ref{fig_correlation} (from \cite{2015MNRAS.453.2576L}) shows the measured correlation pattern of the six best EPTA MSPs as a function of their angular separation. It is clear that the measurement is still uninformative, and no detection can be claimed. Using the same dataset \cite{2015PhRvL.115d1101T} also performed the first search for an anisotropic GWB, constraining any amplitude in the higher multiples of the spherical harmonic decomposition to be less then 40\% of the one in the monopole component (i.e. the isotropic part of the signal). \cite{2016ApJ...821...13A} used the NANOGrav 9yr dataset to place an upper limit of $A=1.5\times10^{-15}$ and to further improve on EPTA limits on the string tension. They also studied spectra with a turnover and placed the first (weak) constraints on the astrophysical properties of the SMBHB population. The best upper limit to date was set to  $A=10^{-15}$ by \cite{2015Sci...349.1522S} using Parkes data. The IPTA \cite{2016MNRAS.458.1267V} also published its first upper limit at $A=1.7\times10^{-15}$ using its first data release. This is not the most stringent limit, but it has been obtained by combining older individual PTA datasets. In fact, compared to the individual dataset used, the IPTA limit is a factor of $\approx2$ better, demonstrating the great potential of adding together more quality pulsars in a worldwide collaborative effort. Figure \ref{fig_hclimit} shows a comparison of the individual PTA upper limits to current theoretical predictions from \cite{2016MNRAS.463L...6S} assuming circular GW-driven SMBHBs, highlighting their uncertainty. In each panel, shaded areas represent the spread (larger than an order of magnitude) due to our poor knowledge of the cosmic galaxy merger rate. The difference between panels stems from different SMBH-galaxy relations (see \cite{2016MNRAS.463L...6S} for model details), highlighting that even a single ingredient entering the computation can have a strong effect on the predicted signal. Moreover, note that possible SMBHB stalling \cite{2016ApJ...826...11S,2017MNRAS.464.3131K}, coupling with the environment \cite{2011MNRAS.411.1467K,2014MNRAS.442...56R} and eccentricity \cite{2007PThPh.117..241E,2013CQGra..30v4014S,2015PhRvD..92f3010H,2016arXiv160607484R} can all contribute to suppressing the signal at low frequencies, which makes any strong inference from GWB non-detection at a $A=10^{-15}$ level problematic.

\section{Future prospects}
\label{sec:future}
In section \ref{sec:tests} and \ref{sec:gws}, we extensively described the state of the art in using pulsars as gravity probes. Precise timing of MSPs in particular, has already provided precise tests of GR and alternative theories of gravity in the strong-field regime, and the current PTA efforts are starting to place constraints on the cosmic population of SMBHBs, although no detection has yet been made. We now take a look at the near future, touching on several subjects that are relevant to improving the use of pulsars as tools for the study of gravity. With pulsars (at least for those that are not dominated by red timing noise, which is the case for most MSPs), the longer the data span, the more precise the TOAs. Therefore simply continuing to monitor them using the same radio telescopes will lead to more precise TOAs. This will in turn lead to more precise estimations of PK and PPN parameters for known pulsar binaries, and therefore more stringent tests of GR and alternative gravity theories. However the number of compact binaries that are useful for gravity tests (for which more than two PK parameters are determined) is limited. More precise TOAs will also naturally improve the constraints on a background of nanohertz GWs (see scaling relations in section \ref{sec:SNR}). However, to make a confident detection of a stochastic GWB, more pulsars with high precision are needed. The search for new pulsars is therefore critical to both tests of strong gravity and the search for nanohertz GWs. 

Next-generation instruments such as the Large European Array for Pulsars (LEAP) (which is equivalent to a 200-m dish and has the sensitivity of SKA phase 1) \cite{2016MNRAS.456.2196B} already play a major role in improving the timing precision of known pulsars and the search for new ones. New instruments such as the South African MeerKAT \cite{2009arXiv0910.2935B} radio telescope are especially promising. With the Five hundred meter Aperture Spherical Telescope (FAST) \cite{2011IJMPD..20..989N} and the Square Kilometre Array (SKA)\cite{2009IEEEP..97.1482D} telescopes, we will take a giant leap in sensitivity, providing both higher precision pulsar timing and surveys that will be able to find thousands of new pulsars \cite{2009A&A...505..919S,2009A&A...493.1161S}. In particular, with its higher sensitivity, we expect major new scientific accomplishments with the arrival of the SKA.

\subsection{High precision timing}

The SKA (SKA1-MID and SKA2) will greatly improve the timing precision of known MSPs, including the Double Pulsar J0737-3039 \cite{bdp+03,lbk+04} and triple systems such as PSR J0337+1715 \cite{rsa+14}. The sensitivity of SKA1-MID should be comparable to that of both Arecibo (with an illuminated surface of about 200 m) and LEAP, but Arecibo has a limited range of observable declinations. For pulsars not visible with Arecibo, the increase in sensitivity will be remarkable. We expect SKA1-MID to improve the pulsar timing precision by one order of magnitude, while SKA2 should improve it by two orders of magnitude \cite{shao2014}. This will provide PTAs with better TOAs and better constraints on a background of GWs, or better, the direct detection of a background or continuous source of GWs. This will also allow us to probe gravity in the strong-field regime, especially with the Double Pulsar PSR J0737-3039, which is visible from the southern sky and is particularly important for SKA. The TOA precision on the Double Pulsar is expected to reach 5 microseconds with SKA1-MID, while reaching sub-microsecond levels with SKA2, which will provide much tighter GR constraints. The eclipse of psr B (with its rapidly-changing flux density) will also be better measured by the SKA. Finally, the moment of inertia of psr A could also be measured and help constrain the equation of state of nuclear matter \cite{watts_ska}. 

\subsection{Finding more MSPs}

The {\it SKA Galactic Census} will enable a search for new pulsars, with which we expect to expand the current pulsar population by a factor of three \cite{keane2014}. In its first phase (SKA phase 1 or SKA1, which includes SKA1 LOW and SKA1 MID), it will have half of the total collecting area expected for SKA2. With its higher sensitivity, the telescope will take less time to achieve a particular S/N, therefore smearing due to varying accelerations will be less of an issue. We expect to find 10,000 normal pulsars with SKA2, including 1800 MSPs \cite{shao2014}. The discovery of new, high-precision MSPs (whether solitary pulsars or binaries) will improve the PTAs' chances of detecting a background of nanohertz GWs or single continuous sources.

\subsection{Finding more highly-relativistic pulsars}

The discovery of new compact binaries such as DNS binaries will lead to more stringent tests of GR \cite{k+06, kbc+04}, while the discovery of new PSR - WD binaries will lead to more stringent tests of alternative theories of gravity, which usually predict the existence of gravitational dipole radiation, a varying gravitational constant and SEP violations \cite{rsa+14}. In particular, we expect to find 100 DNS from SKA1 and 180 DNS from SKA2 \cite{keane2014}. The SKA could also discover new triple systems (a pulsar with two compact objects) that can better constrain the SEP (specifically the $\Delta$ parameter). With a (NS-WD) inner binary and a NS as the outer star, the constraint on $\Delta$ would be many orders of magnitude larger than with PSR J0337+1715 \cite{rsa+14} whose outer star is a WD. Additionally, the improved constraints on PK parameters thanks to the long-term monitoring of known pulsars such as the Double Pulsar, or to new pulsar discoveries could lead us to measure effects at 2PN, which could help constrain the equation of state of neutron stars \cite{watts_ska}. 
The SKA will also improve constraints on the Lense-Thirring effect. Discoveries with the SKA of new DNS binaries could enable a direct measurement of $\dot{x}$ as well as the Lense-Thirring contribution to $\dot{\omega}$ and their time derivatives $\ddot{\omega}$ and $\ddot{x}$. This is especially true for DNS with good timing precision, a close orbit, and a large angle between angular orbital momentum and pulsar spin \cite{shao2014}. 


\subsection{Pulsar - black hole systems}

The discovery of pulsar - black hole (PSR-BH) binaries will open a completely new window on tests of strong gravity \cite{de98,wk99}. Indeed, possible PSR-BH systems are considered to be the {\it holy grail} for testing gravity in the strong-field regime, and better understanding the nature of black holes and their environments: the gravitational potential, the spacetime curvature, the compactness of the black hole. Pulsar timing allows for unique, high precision tests on the nature of black holes 
, including measuring the mass $M_{\rm BH}$, angular momentum $S_{\rm BH}$ and charge $Q_{\rm BH}$ of the black hole. This in turn allows us to test of the {\it cosmic censorship conjecture} (which states that there cannot exist naked singularities) and the {\it no hair theorem} (stating that black holes are completely determined by $M_{\rm BH}$, $S_{\rm BH}$ and $Q_{\rm BH}$), which is violated in some alternative theories of gravity \cite{hp70,kbc+04, liu+12, liu+14,wex+13}.

We expect to find the first PSR-BH binaries with SKA2 \cite{shao2014,liu+14}. There are three types of black hole environments that can be tested: pulsars could be found orbiting a stellar mass BH \cite{ppr05}; an Intermediate Mass Black Hole (IMBH) in globular clusters \cite{dcmp07, csc+14, hessels2014}; a supermassive BH such as Sgr A* \cite{d10}. 
In fact, dozens of radio pulsars are expected to be found in the Galactic Centre orbiting SgrA*. If the pulsars are close enough to the Galactic centre, the spacetime around the supermassive black hole (SgrA*) could be tested \cite{pl+2004}. The magnetar J1745-2900, detected in both X-rays and radio, was found to be near SgrA* \cite{mori2013, kennea+13, eatough+13}. However it is not close enough to it to perform strong gravity tests. If pulsars can be found close enough to SgrA*, the determination of PK parameters through pulsar timing can help find a precise mass of SgrA* (through a measurement of $\dot{\omega}$, $r$, $s$ and $\gamma$). Additionally, the determination of the spin-orbit coupling 
could help place an upper limit on the BH spin. We note that a recent paper suggested that, after analysing decades of pulsar timing data, a known MSP (PSR B1820-30A) is found to be orbiting an IMBH in the globular cluster NGC6624 \cite{perera+17}. While a great result in demonstrating the existence of IMBHs, the orbital period of the system is too long to allow for tests of the spacetime around the black hole. Finally, it will be possible to probe the spacetime around black holes by combining pulsar timing data with the study of relativistic stellar orbits and the high-resolution imaging of the black hole horizon at sub-mm wavelengths, in particular with the Event Horizon Telescope. This will provide a highly-precise test of the no-hair theorem \cite{psaltis2016,EHT}. \subsection{Multi-messenger astronomy}

The LIGO detections of GWs from pairs of black holes were truly historic and provide a new window on tests of strong gravity \cite{LIGO}. In the near future, we expect LIGO and VIRGO to find other sources of GW, such as neutron stars in binary systems (DNS binaries or NS-BH systems) \cite{2010CQGra..27q3001A}, while we expect the space interferometer LISA \cite{2017arXiv170200786A} to find DNS and NS-BH systems with orbital periods of hours to minutes (see, e.g. \cite{2001A&A...375..890N}). The era of multi-messenger astronomy is here: if the neutron star in the binary system is visible as a radio pulsar, the information from GW ground detectors can be combined with the electromagnetic radiation from the pulsar (through pulsar timing and the determination of PK parameters). This will enable us to learn about the environment of GW sources such as neutron stars in binary systems, which emit both radio waves (if the neutron star is visible as a radio pulsar) and GW waves. We might discover new physics, or at the very least, we will better constrain alternative theories of gravity \cite{shao2017}.

\subsection{Detecting nanohertz GWs}

As already mentioned, in the FAST and SKA era, better timing together with the addition of new MSPs to PTAs will greatly increase the sensitivity of PTAs to a background of nanohertz GWs. The use of wideband receivers will help greatly in mitigating all chromatic effects related to pulse propagation in the ISM, allowing a better identification and removal of scattering and dispersion effects. Conversely, achromatic noise sources such as jitter and spin noise cannot be mitigated by extending the receiver band and might eventually dominate the rms uncertainty of the best timed MSPs \cite{2013CQGra..30v4002C}. A promising avenue for improving PTA capabilities is to abandon the idea of analysing pre-determined TOAs, constructed by matching the profiles of individual measurements to a template, and instead directly use the pulse profiles of each individual observation. This type of profile domain analysis has been shown to potentially provide significant improvements in identifying scattering \cite{2017MNRAS.468.1474L} and spin noise \cite{2016ApJ...828L...1S}, and thanks to continuously improving algorithms, it can be also applied to wideband data \cite{2017MNRAS.466.3706L}.

Despite the promise of great improvements, it is extremely difficult to forecast when PTAs will detect GWs. Theoretical exercises \cite{2013CQGra..30v4015S,2016ApJ...819L...6T,2016PhRvD..94l3003V,2017arXiv170202180K} tend to predict a first detection within a decade. This depends on many unknowns, such as how many pulsars will be discovered and added to the PTAs each year, their intrinsic properties and stability, an optimization of the observing schedule (see e.g. \cite{2012MNRAS.423.2642L}), and on the very uncertain GW predictions (cf figure \ref{fig_hclimit}). It is likely that the first glimpse of GWs in the data will come from the stochastic GWB rather than a deterministic source \cite{2015MNRAS.451.2417R}. Given that current PTA sensitivities are dominated by a few, very stable systems, it is likely that this will not significantly change in the near future. The GWB will then show as additional red noise in the best timed pulsars, which will cause a saturation of PTA upper limits before any confident claim can be made through the detection of the HD correlation pattern.

Looking further ahead well within the SKA era, GW detection with PTAs will enable a series of scientific breakthrough including: i) proving the existence of sub-pc SMBHBs, ii) understanding of the dynamics and cosmic history of SMBHBs, ii) identification and sky localization of individual sources, thus enabling multimessenger astronomy of massive objects in the low frequency regime, iii) tests of the existence of extra GW polarizations, iv) unprecedented constraints (or detection) on cosmic (super)string theories and other cosmological sources of GWs, such as first order phase transitions. A useful summary of the GW science enabled by future PTA detections can be found in \cite{2015aska.confE..37J}. 

\section{Summary}
\label{sec:sum}

MSPs are extremely precise clocks. Monitoring the TOAs of their radio pulses over many years (through the process of ``pulsar timing") makes them rival atomic clocks on Earth. This process allows us to determine pulsar parameters with extremely high precision, including their rotation period, period derivative or the dispersion measure due to the interstellar medium. Through this process, we can learn about the strong-gravity environment around the neutron star, the interstellar medium, and possibly detect GWs from distant SMBHBs.  For pulsars in binaries, we can constrain the strong-gravity environment around the neutron star through the fitting of PK parameters. The Double Pulsar PSR J0737-3039 provides the best test so far of GR, confirming GR within 0.05\%. Constraints on violations of the Strong Equivalence Principle (SEP) -- which would signal a breakdown of GR -- as well as constraints on alternative theories of gravity, are however better achieved with pulsar - white dwarf (PSR-WD) binaries such PSR J1738+0333 and PSR J0348+0432. We also expect the triple system PSR J0337+1715 (one PSR with two WD) and future PSR-BH binaries to impose even stronger constraints. TOA precision naturally improves with longer datasets, therefore the long-term monitoring of known pulsars will yield better gravity constraints. However the timing precision of known pulsars such as the Double Pulsar will be largely improved with newer telescopes such as FAST or the SKA, which have much greater sensitivity. The number of interesting pulsar binaries that can constrain gravity theories is small (more than two PK parameters need to be determined in order to provide an extra consistency check on the theory of gravity). Therefore gravity constraints will be much improved with new pulsar searches such as with the SKA: an additional 1800 MSPs are expected to be found, and 10\% of these could be pulsar binaries. In particular, we hope to find dozens of new DNS, which would lead to tighter constraints on GR, while PSR-WD binaries would provide interesting tests on alternative theories of gravity. Additionally, triple systems such as (PSR-WD-WD) systems but also (PSR-WD-NS) would be particularly interesting for constraining the SEP and alternative theories of gravity. The holy grail however would be a PSR-BH system, which would not only provide very tight constraints on gravity theories, but also uniquely probe the spacetime around black holes. Pulsar timing techniques can also be combined with other techniques to provide even more stringent tests: the no-hair theorem can be better constrained by combining data from pulsar timing with the imaging of the BH event horizon using the Event Horizon Telescope \cite{psaltis2016}. In addition, pulsar timing data from pulsar binaries can be combined with future LIGO or LISA GW detections of neutron stars in binary systems, which may lead to the discovery of new physics or impose strong constraints on alternative theories of gravity \cite{shao2017}. Cross-correlating the TOAs from an ensemble of MSPs thus forming a PTA, offers the unique possibility of detecting GWs in the nanohertz frequency range, which is inaccessible to both current ground and future space-based interferometers. The most promising GW sources in this frequency range are a cosmic population of SMBHBs, and the incoherent superposition of their signals will most likely be detected as a stochastic GWB with a red spectrum. Current PTAs already placed stringent limits on the amplitude of such a GWB, skimming the level at which theoretical predictions place the signal. The discovery of new MSPs with the SKA will greatly improve the sensitivity of current PTAs, most likely leading to a confident detection by the end of next decade. Together with LIGO/Virgo/Kagra in the kilohertz and LISA in the millihertz frequency range, PTAs will contribute to making gravitational wave astronomy possible across twelve orders of magnitude in frequency, which will allow us to probe the astrophysics of compact objects and strong gravity across ten orders of magnitudes in mass scale. With the prospects of great improvements ahead and the enormous potential for new discoveries that come with it, pulsar timing stands as a unique tool in the quest to understand the nature of gravity and the Cosmos.

\section{Acknowledgements}
The authors wish to thank Michael Kramer and Norbert Wex for discussions and comments and Sarah Burke-Spolaor for providing figure 3. A.S. is supported by the Royal Society.



\bibliographystyle{ws-rv-van}
\bibliography{bibnote}


\end{document}